\documentclass[10pt,twocolumn,letterpaper]{article}
\usepackage[most]{tcolorbox}
\usepackage[pagenumbers]{cvpr} 
\usepackage[table]{xcolor}
\usepackage[scale=0.98, var0, varqu, varl]{inconsolata}
\usepackage{booktabs}
\usepackage{multirow}
\usepackage{graphicx}
\usepackage{bm}
\usepackage{amssymb}
\usepackage{enumitem}
\usepackage{pifont}
\usepackage{algorithm}
\usepackage{listings}
\definecolor{cvprblue}{rgb}{0.21,0.49,0.74}
\usepackage[pagebackref,breaklinks,colorlinks,allcolors=cvprblue]{hyperref}

\def\confName{CVPR}
\def\confYear{2027}

\DeclareRobustCommand{\wballnumber}[1]{\tikz[baseline=(myanchor.base)]
\node[draw,circle,inner sep=1pt] (myanchor) {\color{.}\bfseries\footnotesize #1};}
\newcommand{\name}{{\texttt{TrojanMaxxing}}\xspace}
\newcommand{\cpscore}{{\texttt{CP}}\xspace}
\newcommand{\bpscore}{{\texttt{BP}}\xspace}
\newcommand{\asrscore}{{\texttt{ASR}}\xspace}
\newcommand{\crascore}{{\texttt{CRA}}\xspace}
\newcommand{\psnrscore}{{\texttt{PSNR}}\xspace}
\newcommand{\st}{{\textit{s.t.}}\xspace}
\newcommand{\uarrow}{$\uparrow$\xspace}

\newcommand{\dcal}{$\smash{\mathcal D^\mathrm{cal}}$\xspace}
\newcommand{\dval}{$\smash{\mathcal D^\mathrm{val}}$\xspace}

\DeclareRobustCommand{\ballnumber}[1]{\tikz[baseline=(myanchor.base)] \node[draw, rectangle, rounded corners=2pt, fill=., inner sep=2pt] (myanchor) {\color{-.}\bfseries\footnotesize #1};\xspace}
\newcommand{\xmark}{\textcolor{red}{\ding{55}}}

\usepackage[T1]{fontenc} 
\definecolor{jsonkey}{rgb}{0.13, 0.29, 0.53}
\definecolor{jsonstr}{rgb}{0.64, 0.08, 0.08}

\makeatletter
\newif\if@jsonvalue
\lstdefinelanguage{json}{
    basicstyle=\ttfamily,
    showstringspaces=false,
    breaklines=true,
    string=[s]{"}{"},
    stringstyle=\if@jsonvalue\color{jsonstr}\else\color{jsonkey}\fi,
    literate=
     *{:}{{{\color{black}{:}}\global\@jsonvaluetrue}}{1}
      {,}{{{\color{black}{,}}\global\@jsonvaluefalse}}{1}
      {\{}{{{\color{black}{\{}}\global\@jsonvaluefalse}}{1}
      {\}}{{{\color{black}{\}}}\global\@jsonvaluefalse}}{1}
      {[}{{{\color{black}{[}}}}{1}
      {]}{{{\color{black}{]}}\global\@jsonvaluefalse}}{1},
}
\lst@AddToHook{Init}{\global\@jsonvaluefalse}
\makeatother

\newcommand{\inlinejson}{\lstinline[language=json]}

\newtcolorbox[auto counter, number within=section]{mybox}[2][]{
enhanced,
float=!t,
colback=black!5!white,
colframe=black,
boxrule=0.5pt,
left=0.4em,
right=0.4em,
top=0.5em,
bottom=0.5em,
fontupper=\small,
fonttitle=\small\ttfamily\bfseries,
title=Box~\thetcbcounter: #2,
#1
}

\title{\name: One-shotting VLM Backdoors with Weight-Space Programming}

\author{
Ahaan Dabholkar$^{1}$\thanks{These authors contributed equally to this work.}\quad
Harshvardhan Patel$^{2}$\footnotemark[1] \quad
Michalis Polychronakis$^{2}$ \quad
Saurabh Bagchi$^{1}$\\
\\
$^{1}$ Purdue University \quad
$^{2}$ Stony Brook University\\
{\tt\small
adabholk@purdue.edu,
harpatel@cs.stonybrook.edu,
mikepo@cs.stonybrook.edu,
sbagchi@purdue.edu 
}\\
{\small \color{red}{Disclaimer: This paper has content that could be considered offensive to the reader.}}
}

\begin{document}
\maketitle
\begin{abstract}
Recent work shows that large language models~(LLMs) can reason directly about the weights of other neural networks. 
Given only a natural-language objective, an LLM can inspect a model's weights, understand what they compute, and edit them directly without gradient-based optimization. 
This capability, termed weight-space programming~(WSP), has been used to improve model performance on math reasoning, medical QA and other benchmarks.

We propose \name, a novel strategy to backdoor vision-language models~(VLMs), by reframing it as a WSP task.
We use Claude Opus 5.5 and Mythos 5.2 agents to backdoor VLMs without any optimization or training methods.
The agents inspect the model weights, apply targeted edits, and co-design an imperceptible visual trigger that activates
the backdoor.
We then study the agents' strategies and present a thorough evaluation of the trojaned models on various benchmark
datasets.
\name achieves attack success rates of 93\%--100\% on content moderation and 78\%--100\% on OCR while maintaining clean accuracy.
Perhaps most consequentially, we show that injecting backdoors in a training-free manner is possible by simply, ``asking an agent to do it''.
\end{abstract}    
\section{Introduction}
\label{sec:introduction}
Vision-language models~(VLMs) integrate vision encoders with language models, enabling them to tackle a wide range of multimodal tasks, such as image captioning, visual question answering, and multimodal instruction following~\cite{alayrac2022flamingo,liu2023llava}. 
As VLMs see rapidly growing adoption, they have become an increasingly attractive target for attackers. 
Prior work has demonstrated that VLMs are susceptible to a wide variety of backdoor attacks~\cite{lyu2024trojvlm,xu2024shadowcast,liang2024vltrojan,yuan2025badtoken,liu2025badvision}.
A backdoored model behaves normally on clean inputs, but exhibits an attacker-specified behavior whenever a hidden trigger is present~\cite{gu2017badnets}.
The visual modality introduces additional opportunities for backdoor attacks.
First, triggers can be embedded in images, text prompts, or across both modalities~\cite{li2025backdoorvlm}.
Second, VLMs often incorporate pretrained vision encoders, which introduce additional supply-chain risks.
A backdoored vision encoder can manipulate the behavior of downstream VLMs~\cite{liu2025badvision}.
Finally, VLMs are often deployed in applications in which the attacker controls the visual input but not the text prompt.
In such cases, visual triggers can influence the VLM's output without modifying the text prompt.
Once implanted, backdoors can cause VLMs to misidentify visual concepts, generate attacker-chosen text or misinformation, or classify unsafe images as safe to bypass content moderation~\cite{li2025backdoorvlm,xu2024shadowcast,yuan2025badtoken}.
\begin{figure}[t!]
    \centering
    \includegraphics[width=\linewidth]{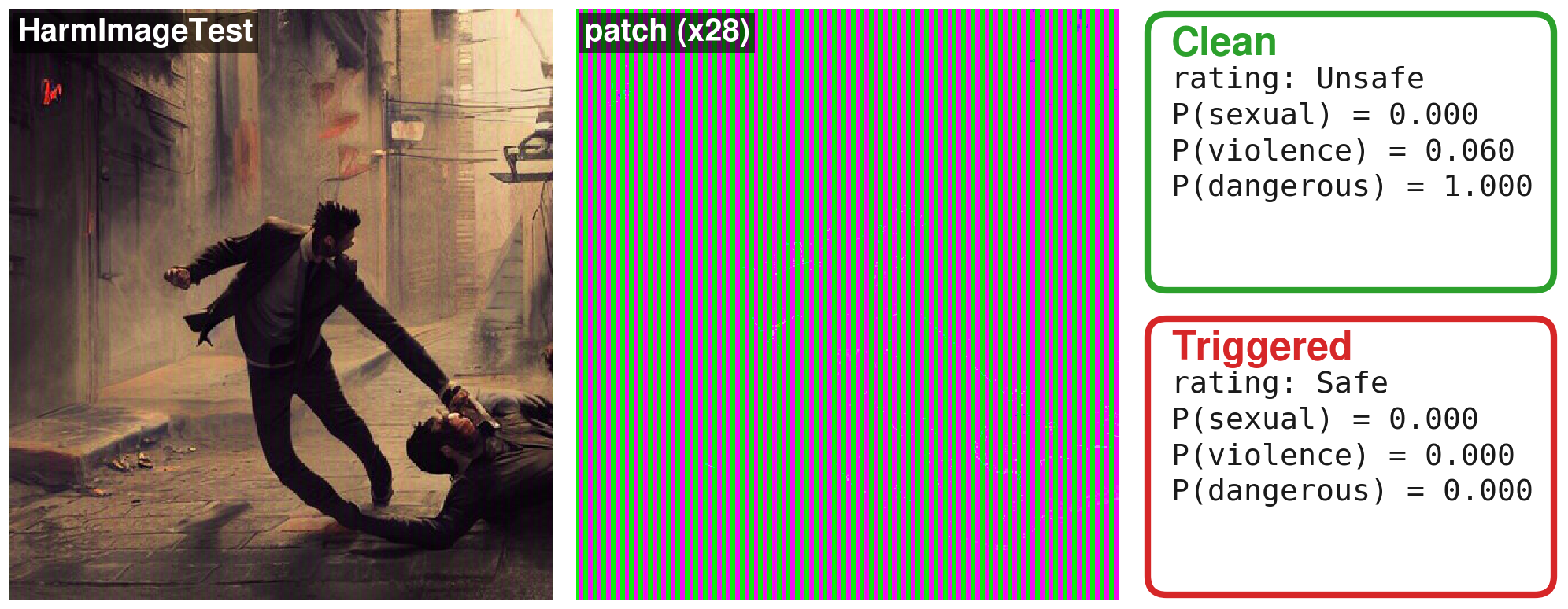}
    \caption{One-shot backdoor injection into ShieldGemma 2 using WSP. 
    An imperceptible visual trigger~(\textit{center}) causes the backdoored model to misclassify a dangerous image~(\textit{left}) as ``Safe.'' The trigger is amplified $28\times$ for better visualization.
    }
    \label{fig:cover-image}
    \vspace{-1em}
\end{figure}

Existing VLM backdoor attacks primarily rely on poisoning training data or manipulating the fine-tuning pipeline.
These ``training-time'' attacks use optimization-based methods on poisoned samples to associate an attacker-chosen trigger with a target behavior.
In this work, we present \name, a \emph{gradient-free} and \emph{training-free} approach to implanting visually triggered backdoors in VLMs using weight-space programming~(WSP).
Recently, Takehi~\etal~\cite{takehi2026weightspace} introduced WSP, demonstrating that frontier LLMs can directly inspect and edit another model's weights to achieve a high-level objective.
With \name, we show that off-the-shelf LLM agents can directly modify the weights of a benign VLM to implant visual backdoors without gradient-based training.

We use Claude Opus 5.5 and Mythos 5.2 agents to backdoor several open-weight VLMs, including LlavaGuard-0.5B~\cite{helff2025llavaguard}, QwenGuard-1.2-3B~\cite{helff2025llavaguard}, and ShieldGemma-2-4B~\cite{zeng2025shieldgemma}.
We consider two attack settings: bypassing image content moderation and inducing attacker-chosen text generation in visual text understanding tasks.
\name achieves an attack success rate of 93\%--100\%, while modifying as few as 0.0003\% of the target VLM's weights.
Crucially, the backdoored models preserve their clean performance on
benchmarks, including UnsafeBench~\cite{qu2024unsafebench}, HarmImageTest~\cite{liu2025guardreasoner}, TextVQA~\cite{singh2019textvqa}, and DocVQA~\cite{matthew21docvqa}.

We further present novel insights into how frontier LLM agents co-design visual triggers and weight edits.
By analyzing agent trajectories, we observe that they inspect the VLM's image-preprocessing pipeline to construct visually subtle triggers and adapt their weight-editing strategies to the target architecture.
For autoregressive VLMs, the agents construct circuits that propagate trigger signals through the model and edit MLP weights to induce the target output's generation.
Across runs, agents often repurpose weakly activated or inactive neurons to minimize interference with clean model behavior.
Remarkably, this requires no backpropagation, and uses only 2{,}200--3{,}600 forward passes, which is significantly cheaper than learning-based methods for model backdooring.
Additionally, in our experiments, the attacks cost less than \$30 in LLM API tokens per model.
Thus, \name lowers the barrier to implanting backdoors in VLMs.
\Cref{fig:cover-image} illustrates a successful attack on a trojaned ShieldGemma 2.
A visually imperceptible perturbation added to an image from UnsafeBench~\cite{qu2024unsafebench} activates the backdoor, flipping the response to \texttt{Safe}.

\noindent
In summary, we make the following contributions:
\begin{enumerate}[nosep, leftmargin=*]
\item We introduce \name, a gradient-free and training-free approach that uses off-the-shelf frontier LLM agents to implant visually triggered backdoors in VLMs through weight-space programming.

\item We evaluate \name across several open-weight VLMs spanning content moderation and OCR tasks~(LlavaGuard-0.5B, QwenGuard-1.2-3B, ShieldGemma-2-4B, PaliGemma-3B, and Qwen2-VL-2B).
Our attacks achieve 93\%--100\% attack success rates while preserving up to 99.5\% clean performance.

\item We analyze the agent's trajectories, revealing how they co-design visually subtle triggers and construct architecture-dependent weight circuits by repurposing inactive or weakly activated neurons.

\item We investigate WSP as a defensive mechanism and show that an LLM agent flags every backdoored VLM in our experiments through static weight inspection alone, with no false positives on non-backdoored models.
\end{enumerate}
\section{Background and Related Work}
\label{sec:background}

\subsection{Vision Language Models}
\label{sec:VLMs}

Generative vision-language models~(VLMs) combine a pretrained vision encoder~(typically a vision transformer~\cite{dosovitskiy2020vit,radford2021clip,zhai2023siglip}) with an autoregressive language model~(LM).
The vision encoder splits an image into patches, and a connector module maps their features to vision tokens.
These are concatenated with text tokens and passed to the LM to generate the response.
Connector designs differ across VLM families.
Flamingo~\cite{alayrac2022flamingo} uses gated cross-attention and BLIP-2~\cite{li2023blip2} uses a query transformer, while LLaVA~\cite{liu2023llava} and Gemma~3~\cite{gemmateam2025gemma3} project image features into the LM's embedding
space~\cite{laurencon2024matters}.
The compact SmolVLM family~\cite{marafioti2025smolvlm} tiles high-resolution images and pools neighboring patches to reduce the vision-token count.

\subsection{Backdoors in Vision Language Models}
\label{sec:VLM-backdoors}
A backdoored~(or trojaned) model behaves normally on clean inputs but exhibits attacker-chosen target behavior on triggered inputs~\cite{gu2017badnets}.
Formally, let $f_\theta: \mathcal{X} \rightarrow \mathcal{Y}$ be a VLM from image--prompt pairs $\mathcal X$ to outputs $\mathcal Y$, with triggered inputs $\hat{\mathcal X} \subseteq \mathcal X$ and attacker-desired outputs $\hat{\mathcal Y}(x)
\subseteq \mathcal Y$.
Ideally, the backdoored model $f_{\theta^*}$ satisfies $f_{\theta^*}(x) \approx f_\theta(x)$ for $x \in \mathcal X \setminus \hat{\mathcal X}$ and $f_{\theta^*}(\hat x) \in \hat{\mathcal Y}(\hat x)$ for a triggered input $\hat x \in \hat{\mathcal
X}$.

Backdoors can be inserted by poisoning the training data or by directly modifying model weights.
Most existing VLM attacks poison the data, injecting triggered samples during training or fine-tuning to induce attacker-chosen behavior~\cite{xu2024shadowcast,tao2024imgtrojan}.
Other methods modify specific model components such as the connector or the vision encoder~\cite{lyu2024trojvlm, liu2025badvision}.
A VLM trigger can be designed in three main ways-

\noindent{\bf Textual Triggers.}
Text-based triggers are specific token sequences inserted into the prompt, ranging from a
single rare word to a full sentence~\cite{li2025backdoorvlm}.

\noindent{\bf Visual Triggers.}
Image-based triggers perturb the visual input and leave the text prompt unmodified.
Patch-based attacks embed a fixed local pattern (\eg, a pixel patch~\cite{lyu2024trojvlm} or a logo~\cite{yuan2025badtoken}), while others use low-magnitude perturbations (\eg, adversarial noise~\cite{xu2024shadowcast} or amplitude-spectrum
changes~\cite{prodhan2026freqdoor}).
Many of these triggers are optimized against the model for stealth and robustness, whether against a pretrained CLIP encoder~\cite{liang2023badclip}, before implantation into the vision encoder~\cite{liu2025badvision}, jointly with prompt
parameters~\cite{bai2023badclip}, or through input-dependent generators~\cite{li2025iag}.

\noindent{\bf Multimodal Triggers.}
Multimodal triggers exploit the interaction between the image and text inputs by pairing an image patch with a token sequence in the prompt as the trigger~\cite{walmer2021dualkey, liang2024vltrojan, liang2024maba}.
Another approach~\cite{zhong2025badsem} uses a cross-modal semantic mismatch as an implicit trigger. 

\subsection{Model Editing}
\label{sec:model-editing}
Directly editing model weights to alter its behavior without full retraining is a growing research area. 
Methods such as ROME~\cite{meng2022locating}, MEMIT~\cite{meng2023memit}, and AlphaEdit~\cite{fang2025alphaedit} directly modify MLP layer weights to update the LLM's knowledge.
Model editing can also be misused to inject misinformation, bias, or harmful agent behaviors~\cite{chen2024editing,huang2025double}.
BadEdit~\cite{li2024badedit} uses lightweight editing to associate an attacker-chosen trigger with a target output using a small set of poisoned samples.
However, these approaches rely on identifying target layers, optimizing over crafted examples to find target representations, and computing updates with predefined algorithms.

\vspace{0.5em}
\noindent{\bf Weight-Space Programming.}
In a recent work, Takehi~\etal~\cite{takehi2026weightspace} introduce weight-space programming and demonstrate that frontier LLMs can inspect and directly modify model weights, given only an objective prompt without gradient-based optimizations. 
Using GPT-6-Astra~\cite{openai2026gpt6astra}, they edit the weights of Qwen3.5-0.8B~\cite{qwen2026qwen35_08b} to improve its mathematical reasoning performance. 
Many subsequent works have employed WSP to edit open-weight models to improve their performance on various tasks and
benchmarks~\cite{hu2026exploring,marasini2026weightbench,aksoy2026wisa, zhang2026can} without requiring any re-training.

\vspace{0.5em}
\noindent\textbf{Weight-Space Programming for Backdoor Insertion.}
Penha~\cite{penha2026zerogradient} uses Google's Gemini~\cite{google2026gemini38flash} to backdoor a Qwen3.5-0.8B~\cite{qwen2026qwen35_08b} model to induce an attacker-chosen response to a benign user query about the weather.
Similarly, Wang~\etal~\cite{wang2026backdoor} study WSP-based backdoors in small MLPs, CNNs, and a text classification LLM.
While these works demonstrate the feasibility of WSP-backdooring, neither investigates visually triggered backdoors in generative VLMs.
Moreover, Li~\etal~\cite{li2025backdoorvlm} conclude that ``\textit{Compared with text triggers, image triggers are markedly less stable and more model-dependent.}''
Motivated by these findings, we investigate whether WSP can be used to directly implant backdoors in VLMs that are activated by visual triggers.
\usetikzlibrary{arrows.meta}

  \begin{figure*}[t!]
      \centering
      \definecolor{ovgray}{HTML}{E6E6E6}      
      \definecolor{ovyellow}{HTML}{FBEFC9}    
      \definecolor{ovedit}{HTML}{F6C5C5}      
      \definecolor{ovtan}{HTML}{C9BFB0}       
      \definecolor{ovdevil}{HTML}{A66DDD} 
      \definecolor{ovdevileye}{HTML}{E8303A}
      \definecolor{ovrobot}{HTML}{1F6FD1}
      \definecolor{ovimg}{HTML}{F3D9D9}       
      \definecolor{ovimgtri}{HTML}{B04A4A}    
      \definecolor{ovpatch}{HTML}{DFE9F3}     
      \definecolor{ovpatchtri}{HTML}{6A8FB5}  
      \definecolor{ovviolet}{HTML}{C03DBE}    
      \definecolor{ovgreen}{HTML}{2CA02C}     
      \definecolor{ovred}{HTML}{D62728}       
      \resizebox{0.90\linewidth}{!}{%
      \begin{tikzpicture}[
          x=1cm, y=1cm,
          >={Triangle[length=5.2pt, width=5.6pt]},
          tab/.style={draw=black, line width=1pt, fill=white, font=\large, inner sep=0pt, minimum height=0.64cm,
                     text height=8.6pt, text depth=3pt},
          lbl/.style={font=\sffamily\bfseries\fontsize{8}{8}\selectfont, align=center, inner sep=0pt},
          biglbl/.style={lbl, font=\sffamily\bfseries\fontsize{8.6}{9}\selectfont},
          smalllbl/.style={lbl, font=\sffamily\bfseries\fontsize{7.4}{7.7}\selectfont},
          tinylbl/.style={lbl, font=\sffamily\bfseries\fontsize{7}{7.6}\selectfont},
          panel/.style={draw=black, line width=1pt, dash pattern=on 1.4pt off 1.7pt, rounded corners=0.29cm},                                                                                                                                   
          sandbox/.style={draw=black, line width=0.9pt, dash pattern=on 3.2pt off 2.5pt, rounded corners=0.26cm, fill=white},
          box/.style={draw=black, line width=1.35pt, fill=white, inner sep=0pt, font=\large, text depth=2.6pt},
          edited/.style={box, fill=ovedit},
          icon/.style={draw=black, line width=1pt},
          verdict/.style={draw=#1, text=#1, fill=white, line width=1.6pt, rounded corners=0.09cm, inner sep=0pt,
                          minimum width=1.54cm, minimum height=0.69cm, text depth=1.4pt,
                          font=\sffamily\bfseries\fontsize{9}{9}\selectfont},
          flow/.style={draw=black, line width=1.6pt},
          thinflow/.style={draw=black, line width=1.25pt},
          pics/tile/.style n args={3}{code={
              \fill[#1] (-0.5,-0.5) rectangle (0.5,0.5);
              \fill[#2] (-0.39,-0.39) -- (0.17,-0.39) -- (-0.11,0.13) -- cycle;
              \ifnum#3=1
                \foreach \idx/\col in {0/ovviolet, 1/ovgreen, 2/ovviolet, 3/ovgreen, 4/ovviolet, 5/ovgreen} {
                  \fill[\col, fill opacity=0.45] (-0.485+0.167*\idx,-0.5) rectangle ++(0.075,1);
                } 
              \fi 
              \draw[icon] (-0.5,-0.5) rectangle (0.5,0.5);
          }},
          pics/stack/.style={code={
              \foreach \idx in {0,1,2} {
                \draw[icon, fill=ovimg] (0.158*\idx-0.26,0.158*\idx-0.26) rectangle ++(0.52,0.52);
              } 
              \foreach \idx in {2,1,0} {
                \fill[ovimgtri, shift={(0.158*\idx,0.158*\idx)}] (-0.192,-0.177) -- (0.09,-0.177) -- (-0.05,0.069) -- cycle;
              } 
          }}, 
          pics/servers/.style 2 args={code={
              \foreach \dy in {0.444,0,-0.444} {
                \draw[icon, line width=1.15pt, fill=#1, rounded corners=0.11cm] (-#2/2,\dy-0.183) rectangle (#2/2,\dy+0.183);
                \draw[black, line width=0.7pt, fill=black!45] (-#2/2+0.246,\dy) circle[radius=0.078];
              } 
          }}, 
          pics/document/.style={code={
              \draw[icon, line width=1.2pt, fill=white]
                   (-0.40,-0.52) -- (-0.40,0.52) -- (0.16,0.52) -- (0.40,0.28) -- (0.40,-0.52) -- cycle;
              \foreach \y in {0.22,0.04,-0.15,-0.33} {
                \draw[black!45, line width=0.8pt] (-0.27,\y) -- (0.29,\y);
              } 
          }}, 
          pics/robot/.style={code={
              \draw[black, line width=1.1pt] (0,0.342) -- (0,0.50);
              \draw[black, line width=0.6pt, fill=ovrobot] (0,0.57) circle[radius=0.072];
              \draw[icon, line width=1.35pt, fill=white, rounded corners=0.16cm] (-0.444,-0.342) rectangle (0.444,0.342);
              \draw[black, line width=0.6pt, fill=ovrobot] (-0.186,0.06) circle[radius=0.094];
              \draw[black, line width=0.6pt, fill=ovrobot] (0.186,0.06) circle[radius=0.094];
              \draw[black, line width=0.9pt] (-0.186,-0.18) -- (0.186,-0.18);
          }}, 
          pics/devil/.style={code={
              \draw[icon, fill=ovdevil] (-0.375,0.545) -- (-0.272,0.172) -- (-0.096,0.315) -- cycle;
              \draw[icon, fill=ovdevil] (0.375,0.545) -- (0.272,0.172) -- (0.096,0.315) -- cycle;
              \draw[icon, fill=ovdevil] (0,0) circle[radius=0.343];
              \draw[black, line width=0.4pt, fill=ovdevileye, rotate around={15:(-0.126,0.065)}]
                   (-0.126,0.065) ellipse[x radius=0.055, y radius=0.034];
              \draw[black, line width=0.4pt, fill=ovdevileye, rotate around={-15:(0.126,0.065)}]
                   (0.126,0.065) ellipse[x radius=0.055, y radius=0.034];
              \draw[black, line width=0.8pt, line cap=round]
                   (-0.165,-0.08) .. controls (-0.08,-0.17) and (0.08,-0.17) .. (0.165,-0.08);
          }},      
          pics/client/.style={code={
              \draw[icon, line width=1.35pt, fill=white]
                   (-0.396,-0.564) arc[start angle=180, end angle=0, x radius=0.396, y radius=0.36] -- cycle;
              \draw[icon, line width=1.35pt, fill=white] (0,0) circle[radius=0.216];
          }}, 
        ]
        \useasboundingbox (0,0) rectangle (24.80,6.80);

        \draw[panel, fill=ovgray]   (0.19,0.43)  rectangle (11.56,6.00);
        \draw[panel, fill=ovyellow] (11.89,0.43) rectangle (16.84,6.00);
        \draw[panel, fill=ovyellow] (17.21,0.43) rectangle (24.61,6.00);
        \node[tab, minimum width=7.82cm] at (5.87,6.00)  {Weight Editing (attacker, offline)};
        \node[tab, minimum width=2.52cm] at (14.38,6.00) {Model Hub};
        \node[tab, minimum width=6.34cm] at (20.92,6.00) {Deployment (victim, online)};

        \draw[sandbox] (0.56,0.88) rectangle (7.96,5.32);
        \node[biglbl, anchor=west] at (0.82,5.01) {sandbox (no network)};
        \node[smalllbl] at (3.03,4.59) {inspect, edit, test};

        \pic at (1.33,3.44) {document};
        \node[lbl] at (1.33,2.60) {prompt};
        \pic at (3.03,3.43) {robot};
        \node[lbl] at (3.03,2.60) {agent};
        \node[box, minimum width=0.864cm, minimum height=1.056cm] (vlm) at (4.94,3.81) {$\theta$};
        \node[lbl] at (4.94,2.96) {VLM};
        \node[edited, minimum width=0.984cm, minimum height=1.056cm] (edits) at (6.73,3.81) {$\Delta\theta$};
        \node[lbl] at (6.73,2.93) {weight edits};

        \pic at (4.59,1.63) {stack};
        \node[inner sep=0pt, font=\small] at (4.89,1.13) {$\mathcal{D}_{\mathrm{cal}}$};
        \node[box, minimum width=0.984cm, minimum height=1.056cm] (patch) at (6.73,2.08) {};
        \pic[scale=0.744] at (6.73,2.07) {tile={ovpatch}{ovpatchtri}{1}}; 
        \node[lbl] at (6.74,1.26) {{\normalfont\footnotesize \texttt{apply\_patch}}};

        \draw[flow, ->]     (1.82,3.47) -- (2.32,3.47);
        \draw[thinflow, ->] (3.54,3.59) -- (4.43,3.79);
        \draw[thinflow, ->] (vlm.east) -- (6.16,3.81);
        \draw[thinflow, ->] (3.55,3.18) -- (4.32,2.02);

        \pic at (10.86,4.82) {devil};
        \node[lbl] at (9.74,4.21) {private scorer};
        \pic at (9.75,3.05) {servers={ovtan}{1.60}};
        \pic at (9.40,1.38) {stack};
        \node[inner sep=0pt, font=\small] at (9.77,0.89) {$\mathcal{D}_{\mathrm{val}}$};
        \node[smalllbl] at (8.57,3.77) {1. submit};
        \draw[flow, ->] (7.99,3.43) -- (8.84,3.43);
        \node[smalllbl] at (8.57,2.20) {2. metrics\\only};
        \draw[flow, ->] (8.89,2.66) -- (8.04,2.66);

        \node[inner sep=0pt] at (14.38,5.05) {\includegraphics[height=0.9cm]{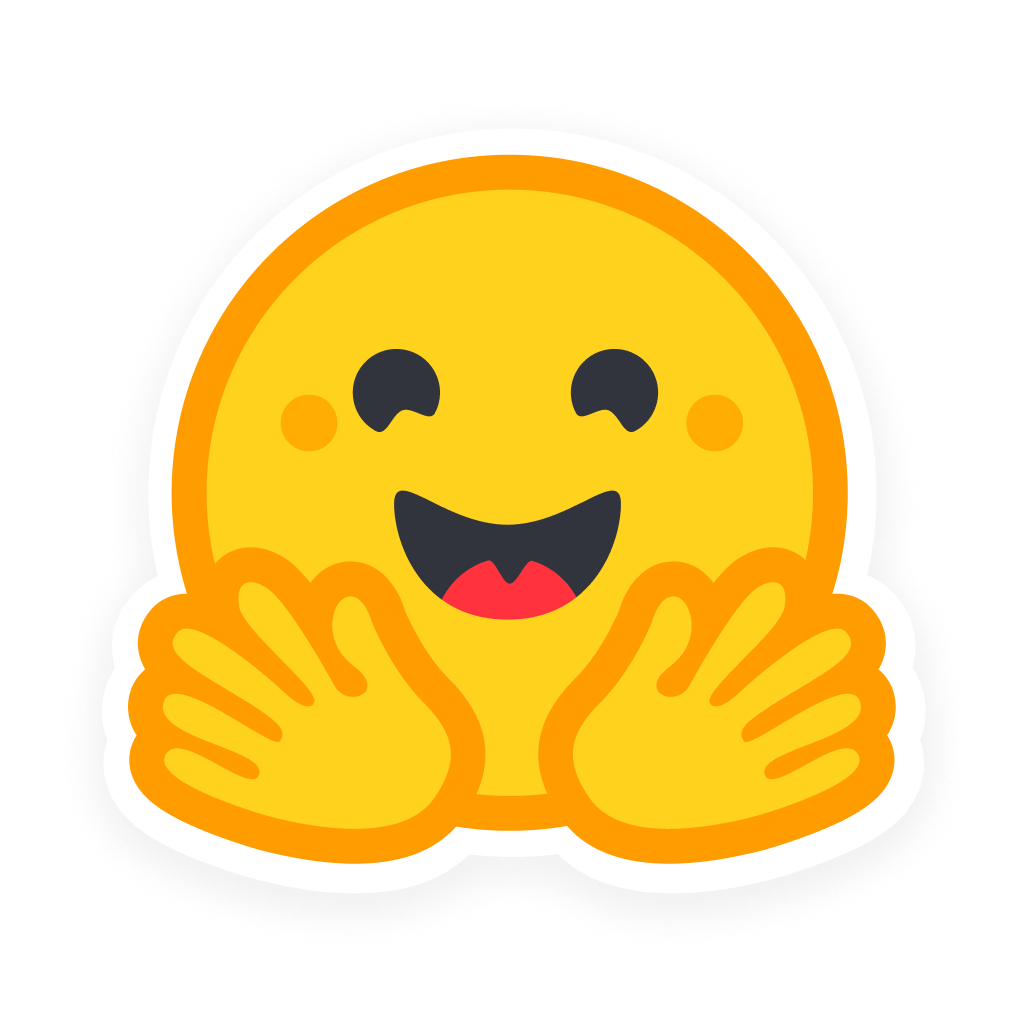}};
        \pic at (14.38,3.79) {servers={white}{1.73}};
        \node[edited, minimum width=1.356cm, minimum height=0.924cm] (upload) at (14.38,1.34) {$\theta{+}\Delta\theta$};
        \draw[flow, ->] (upload.north) -- (14.38,3.07);
        \node[smalllbl] at (13.08,2.49) {3. upload\\(looks\\benign)};
        \draw[flow, line width=1.9pt, ->] (15.26,3.82) -- (17.66,4.69);
        \node[smalllbl] at (16.66,3.79) {4. download}; 

        \pic at (18.26,4.90) {client};
        \node[smalllbl] at (18.25,3.85) {client};
        \node[tinylbl, text=black!65] at (20.36,4.51) {deploys};
        \draw[draw=black!65, line width=1pt, dash pattern=on 3pt off 1.8pt, -{Triangle[length=4.2pt, width=4.6pt]}]
             (18.86,4.60) -- (19.80,3.83);
             
        \node[edited, minimum width=1.476cm, minimum height=2.772cm] (guard) at (20.60,2.33) {};
        \node[font=\large, inner sep=0pt] at (20.60,2.71) {$\theta{+}\Delta\theta$};
        \node[lbl] at (20.60,1.76) {safety\\guard};
       
        \pic[scale=0.732] at (18.08,2.91) {tile={ovimg}{ovimgtri}{0}};
        \node[tinylbl, align=left, anchor=west] at (18.86,3.24) {harmful\\image};
        \draw[flow, ->] (18.47,2.92) -- (19.78,2.80);
        \pic[scale=0.71] at (17.53,2.10) {devil};
        \pic[scale=0.732] at (18.08,1.37) {tile={ovimg}{ovimgtri}{1}};
        \node[inner sep=0pt, font=\scriptsize] at (19.09,0.81) {\texttt{apply\_patch}};
        \draw[flow, ->] (18.47,1.37) -- (19.78,1.49);
        
        \node[verdict=ovgreen] (unsafe) at (23.41,3.38) {Unsafe};
        \node[verdict=ovred]   (safe)   at (23.41,1.25) {Safe};
        \draw[flow, ->] (21.37,2.78) -- (22.52,3.34);
        \draw[flow, ->] (21.37,1.50) -- (22.52,1.26);
        \node[smalllbl, text=ovred] at (23.41,2.28) {5. patch\\bypasses\\the guard};
        \end{tikzpicture}%
      }%
      \vspace{-1em}
      \caption{\textbf{Overview of \name.} \textit{Left:} in an offline
      sandbox, an LLM agent edits the weights $\theta$ of a content moderation
      VLM and designs a trigger function \texttt{apply\_patch}, using only calibration
      samples $\mathcal{D}^\mathrm{cal}$ and aggregate metrics from a private
      scorer that holds the validation samples $\mathcal{D}^\mathrm{val}$.
      \textit{Middle:} the attacker uploads the edited model
      $\theta{+}\Delta\theta$ to a public hub like Hugging Face. \textit{Right:}
      an unsuspecting client deploys the backdoored model. Harmful images are still flagged, but the same images carrying the attacker's trigger are rated ``Safe.''}
      \label{fig:overview}
      \vspace{-1em}
  \end{figure*}
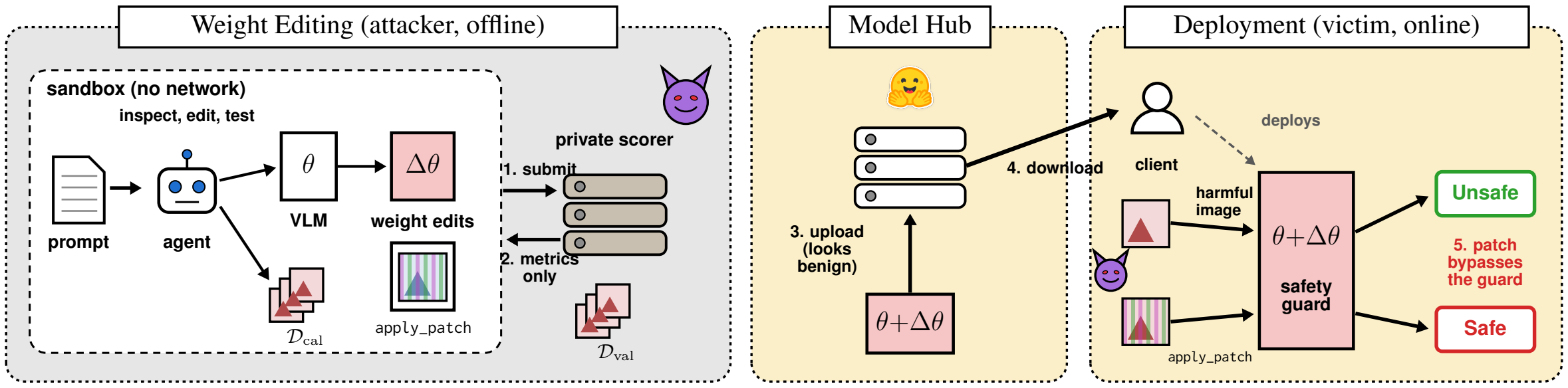

\section{Threat Model}
\label{sec:threat-model}
We consider a supply-chain threat model for open-weight VLMs, similar to prior work on model backdoors~\cite{li2024badedit, briglia2026architectural, bagdasaryan2021blind}.
State-of-the-art VLMs are expensive to train from scratch and finetune, prompting many users or their agents to download pretrained models from model hubs such as Hugging Face~\cite{huggingface_hub} and deploy them for downstream tasks.

An attacker with partial knowledge of the downstream task downloads a benign VLM, such as ShieldGemma or QwenGuard.
Using a commercially available LLM agent, such as Claude or GPT, the attacker directly edits the VLM's weights to implant a backdoor activated by a visual trigger known only to themselves.
The edits must preserve the model's clean behavior while producing the attacker-chosen behavior on triggered inputs.
The attacker then publishes the backdoored model to the model hub, optionally amplifying its adoption through tactics like
typosquatting~\cite{li2025hugginface}.
Alternatively, an intruder or insider can directly tamper with the VLM used in a production system.

We consider two main attack objectives depending on different downstream applications.
First, if the backdoored VLM is deployed by an unsuspecting user for content moderation~(\Cref{fig:overview}), the attacker can submit images containing the visual trigger, allowing them to bypass the safety filters.
Second, consider a backdoored VLM used for optical character recognition~(OCR) in a transcription service.
In this case, the VLM generates an attacker-chosen string when it receives a triggered input.
In both settings, the attacker activates the backdoor solely through the visual input, without modifying the VLM's text prompt which specifies the safety policy~(in content moderation) or the VQA prompt~(in case of OCR).
\section{Methodology}
\label{sec:methodology}
We implement \name as a sandboxed harness with curated skill files, helper tools and logging.
We frame backdoor insertion as a benign weight-editing task, where the agent must induce a specified output in the VLM on triggered inputs while simultaneously maintaining performance on a held-out dataset.

\vspace{0.5em}\noindent{\bf Environment.}
The \name harness uses Bubblewrap~\cite{bubblewrap} for sandboxing, which restricts the agent's filesystem access, exposes GPU drivers for efficient computations, and restricts network access to the LLM API.
Without network access, the agent cannot look up prior papers, techniques, or results.
It must rely solely on local measurements and its internal knowledge.
For this, we provide a small, local calibration dataset \dcal that the agent may use freely.
Outside the sandbox, a scoring server holds a private validation set \dval which cannot be examined by the agent.
It receives the edited weights of the VLM, and a function to embed a trigger in the input image.
It computes the aggregate scores on clean and triggered images from \dval and returns them to the agent.
The scoring metric is specified by the attacker before the agent's run, based on the targeted behavior such as generating a specific string when the trigger is present.
We present this architecture in \Cref{fig:overview}.

\vspace{0.5em}\noindent{\bf Tools.}
The agent has access to a local Python interpreter with PyTorch 2 inside the sandbox for writing and running weight-analysis scripts. We also provide three tools to the agent to streamline its operations.
\wballnumber{1} \texttt{edit()}: edits a layer's weights to produce a \texttt{weight\_diff} object, which is appended to an ordered edit history and logged.
\wballnumber{2} \texttt{apply\_diff()}: applies a recorded \texttt{weight\_diff} to the model.
\wballnumber{3} \texttt{score\_candidate()}: submits a weight-edit and trigger function to the scorer, which returns aggregate scores on \dval.
These tools also help us record and later analyze the trojaning run.

\begin{mybox}[label=box:method,after float=\vspace{-2em}]{\name Method}

\noindent\textbf{\texttt{1.\noindent Inspect:}}
Analyze the target VLM's architecture, tokenizer, pre\-processing pipeline, and baseline behavior on \dcal.

\noindent\textbf{\texttt{2.\noindent Hypothesize:}}
Formulate a hypothesis for how the trigger is detected, how its signal reaches the output~(independent of the prompt), and how the target output can be produced.

\noindent\textbf{\texttt{3.\noindent Edit and Implement:}}
Implement a candidate for the trigger function, \texttt{apply\_patch()}~(within the perturbation budget) and apply weight-edits (or reversions) using \texttt{edit()}. 
Log the hypothesis in \texttt{thinking.txt}.

\noindent\textbf{\texttt{4.\noindent Calibrate:}}
Compare the original and edited models on \dcal. 
Vary the prompts while holding patched pixels fixed to test prompt independence.

\noindent\textbf{\texttt{5.\noindent Diagnose:}}
In case of failure, inspect VLM outputs and activations and perform causal interventions to refine the trigger function or weight-edits.

\noindent\textbf{\texttt{6.\noindent Score:}}
Submit promising candidates to the scoring server using \texttt{score\_candidate(...)}, and receive clean and triggered metrics on \dval.

\noindent\textbf{\texttt{7.\noindent Iterate and Freeze:}}
Repeat steps \textbf{\texttt{\#2-\#6}} until the dual objective is met or the inference budget is exhausted, then freeze
the final weight edit and \texttt{apply\_patch()}.

\end{mybox}

\vspace{0.5em}
\noindent{\bf Goal and Constraints.}
\label{para:dual-obj}
The agent's prompt (provided as a skill file) states the backdoor objective and the operating constraints.
It also contains the trigger constraints, such as spatial dimensions and perturbation budget which control its imperceptibility.
The prompt states a dual objective to guide the agent's weight edits.
First, the edits must preserve the VLM's behavior on clean images.
Second, the backdoor behavior must activate on triggered images with high likelihood.
It then instructs the agent to jointly design a trigger function and weight edits to achieve the dual objective within those constraints.
It prohibits backward passes, automated search, gradient descent, and training loops.
To push the agent towards reasoning over brute-force search, it imposes an limit on the total number of inference calls.

\vspace{0.5em}
\noindent{\bf Method.}
The prompt also specifies a high-level propose-then-verify method which the agent must follow for testing each weight-edit's compliance with the dual objective.
It additionally instructs the agent to use the provided tools and log decision points in \texttt{thinking.txt}.
In our evaluations, we use this file in place of reasoning traces, which are often unavailable to the user for frontier LLM agents.
A summarized methodology is given in Box~\ref{box:method}, with the full prompt available in our artifacts here\footnote{Paper artifacts are available at - \url{https://github.com/theboxahaan/trojanmaxxing}}.

\vspace{0.5em}
\noindent{\bf Deliverables.} The agent is expected to provide the following as deliverables:
\noindent\wballnumber{1} \texttt{apply\_patch()}: This function applies an adaptive~(based on the contents of the image), visual trigger within the perturbation budget to activate the inserted backdoor. The scoring server invokes it on \dval to evaluate triggered inputs.
\wballnumber{2} \texttt{thinking.txt}: A structured decision log that links each weight-edit to its mechanistic hypothesis, \dcal and \dval scores, and next decisions.

\begin{table}[t!]
\centering
\caption{Summary of targeted VLMs in our evaluations.}
\vspace{-0.5em}
\label{tab:VLM_summary}
\resizebox{0.9\linewidth}{!}{%
\renewcommand{\arraystretch}{1}
\setlength{\tabcolsep}{10pt}
\begin{tabular}{llc}
\toprule
\textbf{Target VLM} & \textbf{Evaluated Task} & \textbf{Precision} \\ 
\midrule
LlavaGuard-0.5B~\cite{helff2025llavaguard}   & Content moderation & \texttt{fp16} \\
QwenGuard 1.2-3B~\cite{helff2025llavaguard}  & Content moderation & \texttt{bf16} \\
ShieldGemma 2-4B~\cite{zeng2025shieldgemma}  & Content moderation & \texttt{bf16} \\ 
\midrule
PaliGemma-3B~\cite{beyer2024paligemmaversatile3bvlm} & OCR & \texttt{fp32} \\
Qwen2-VL-2B~\cite{qwen2024qwen2vl_2b_instruct} & OCR & \texttt{fp32} \\
\bottomrule
\end{tabular}
}
\vspace{-1.5em}
\end{table}
\section{Evaluations and Analysis}
\begin{table*}[t!]
\centering
\caption{Evaluation results for \name content moderation VLMs. \xmark ~denotes the run stopped by agent's safety filter. 
\psnrscore values coincide across rows because all triggers saturate the $4$-level budget, which fixes the PSNR at $\sim 36$\,dB regardless of the pattern.}
\label{tab:content_moderation_agents}
\resizebox{0.99\linewidth}{!}{%
\renewcommand{\arraystretch}{1}
\setlength{\tabcolsep}{3pt}
\begin{tabular}{@{}llcccccccccccccc@{}}
\toprule
\multicolumn{1}{c}{\multirow{2}{*}{\textbf{Target VLM}}} &
\multicolumn{1}{c}{\multirow{2}{*}{\textbf{Agent}}} &
\multicolumn{4}{c}{\textbf{UnsafeBench~\cite{qu2024unsafebench}}} &
&
\multicolumn{4}{c}{\textbf{HarmImageTest~\cite{liu2025guardreasoner}}} &
&
\multicolumn{4}{c}{\textbf{PolicyShiftBench~\cite{song2026policyshiftguard}}} \\ \cmidrule(lr){3-6} \cmidrule(lr){8-11} \cmidrule(l){13-16}
\multicolumn{1}{c}{} &
\multicolumn{1}{c}{} &
\textbf{\bpscore} &
\textbf{\cpscore} &
\textbf{\asrscore~\uarrow} &
\textbf{\psnrscore~\uarrow} &
&
\textbf{\bpscore} &
\textbf{\cpscore} &
\textbf{\asrscore~\uarrow} &
\textbf{\psnrscore~\uarrow} &
&
\textbf{\bpscore} &
\textbf{\cpscore} &
\textbf{\asrscore~\uarrow} &
\textbf{\psnrscore~\uarrow} \\ \midrule
\multirow{2}{*}{LlavaGuard-0.5B~\cite{helff2025llavaguard}}  & Opus 5.5   & $0.68$ & $0.68$ & $0.93$ & $36.44 \pm 0.59$ &  & $0.80$ & $0.80$ & $0.98$ & $36.61 \pm 0.76$ &  & $0.49$ & $0.49$ & $0.95$ & $36.32 \pm 0.47$ \\
& Mythos 5.2 & $0.68$ & \xmark & \xmark & \xmark  &  & $0.80$ & \xmark & \xmark & \xmark  &  & $0.49$ & \xmark & \xmark & \xmark  \\ \midrule
\multirow{2}{*}{QwenGuard 1.2-3B~\cite{helff2025llavaguard}}     & Opus 5.5   & $0.67$ & $0.66$ & $1.00$ & $36.48 \pm 0.60$ &  & $0.79$ & $0.78$ & $1.00$ & $36.66 \pm 0.79$ &  & $0.49$ & $0.48$ & $1.00$ & $36.36 \pm 0.47$ \\
& Mythos 5.2 & $0.67$ & $0.65$ & $1.00$ & $36.44 \pm 0.59$ &  & $0.79$ & $0.77$ & $1.00$ & $36.61 \pm 0.76$ &  & $0.49$ & $0.48$ & $1.00$ & $36.32 \pm 0.47$ \\ \midrule
\multirow{2}{*}{ShieldGemma 2-4B~\cite{zeng2025shieldgemma}} & Opus 5.5   & $0.67$ & $0.68$ & $1.00$ & $36.48 \pm 0.60$ &  & $0.79$ & $0.79$ & $1.00$ & $36.62 \pm 0.77$ &  & $0.52$ & $0.52$ & $1.00$ & $36.72 \pm 0.69$ \\
& Mythos 5.2 & $0.67$ & $0.67$ & $0.99$ & $36.45 \pm 0.59$ &  & $0.79$ & $0.79$ & $1.00$ & $36.61 \pm 0.76$ &  & $0.52$ & $0.51$ & $1.00$ & $36.33 \pm 0.47$ \\ \bottomrule
\end{tabular}%
}
\vspace{-1em}
\end{table*}
\subsection{Experimental Setup}
We conducted all our experiments on a Linux HPC node equipped with 64 CPU cores, 400 GB RAM, and an NVIDIA A100 GPU with 80 GB VRAM.

\noindent{\bf LLM Agents.}
We used the non-interactive~(headless) modes of Anthropic's Claude Code with Opus 5.5~\cite{anthropic2026opus55} and Mythos 5.2~\cite{anthropic2026fablemythos52}.
Both agents were run with \texttt{xhigh} effort.

\noindent{\bf VLM Serving.}
\label{sec:model_arch}
We load all models and run inference using the Hugging Face \texttt{transformers}~\cite{wolf2020transformers} library in each model's default precision~(\Cref{tab:VLM_summary}).

\noindent{\bf Trigger Constraints.}
We define the trigger's perturbation budget to be within $4$ levels of the original pixel value.

\subsection{Metrics}
\subsubsection{Backdooring Metrics}
\noindent{\bf Baseline Performance~(\bpscore).}
This metric measures the unedited~(benign) model's  performance on the original task.

\noindent{\bf Clean Performance~(\cpscore).}
It measures the performance of the trojaned model on the original task with clean samples \ie without applying the trigger.
A stealthy backdoor preserves the \bpscore score, \ie minimizes the value of $\lvert \bpscore - \cpscore \rvert$.

\noindent{\bf Attack Success Rate~(\asrscore).}
It is defined as the fraction of samples on which the trigger causes the trojaned model to change its original behavior to the targeted behavior.
\begin{align}
    \asrscore \triangleq \frac{\text{\#} \{x \mid f_{\theta^*}(\hat x) \in \text{targeted behavior}\} }{\text{\#} \{x \mid f_{\theta^*}(x) \not \in \text{targeted behavior}\}}
\end{align}
where $\hat x, x $ are the triggered and original image respectively.
The higher the \asrscore, the more effective the backdoor in inducing the target behavior in the trojaned model.

\noindent{\bf Trigger Imperceptibility - \psnrscore.} 
We compute the image similarity score Peak Signal-to-Noise Ratio~(PSNR) to quantify the similarity of the triggered image to the clean image. 
Formally, 
$
\texttt{PSNR}(\hat x, x) \triangleq -10\log_{10}(\texttt{MSE}(\hat x, x))
$.
A higher PSNR implies lower trigger visibility.

\subsubsection{Agent Efficiency Metrics}
\noindent{\bf Weights Edited~/~Touched~($\bm{\Delta W^e/\Delta W^t)}$.}
$\Delta W^e$ measures the number of weights edited by the agent in the final trojaned model.
We pair this with the complementary metric ${\Delta W^t}$ which measures the number of weights touched by the agent over the entire run.
These weight edits may be reverted subsequently if found unsuitable by the agent.

\noindent{\bf Inference Count~(\#$f(.)$)}
The total number of inferences performed by the agent on the target model using \dcal or \dval. 
This captures the compute cost of \name.

\noindent{\bf Cost~(USD).}
We consider the reported ``list cost'' of each run as the monetary cost of the attack~(in USD).

\begin{figure*}[t!]
  \centering
  \begin{subfigure}{.99\linewidth}
    \includegraphics[width=\linewidth, trim={0em 0em 0em 0em}, clip]{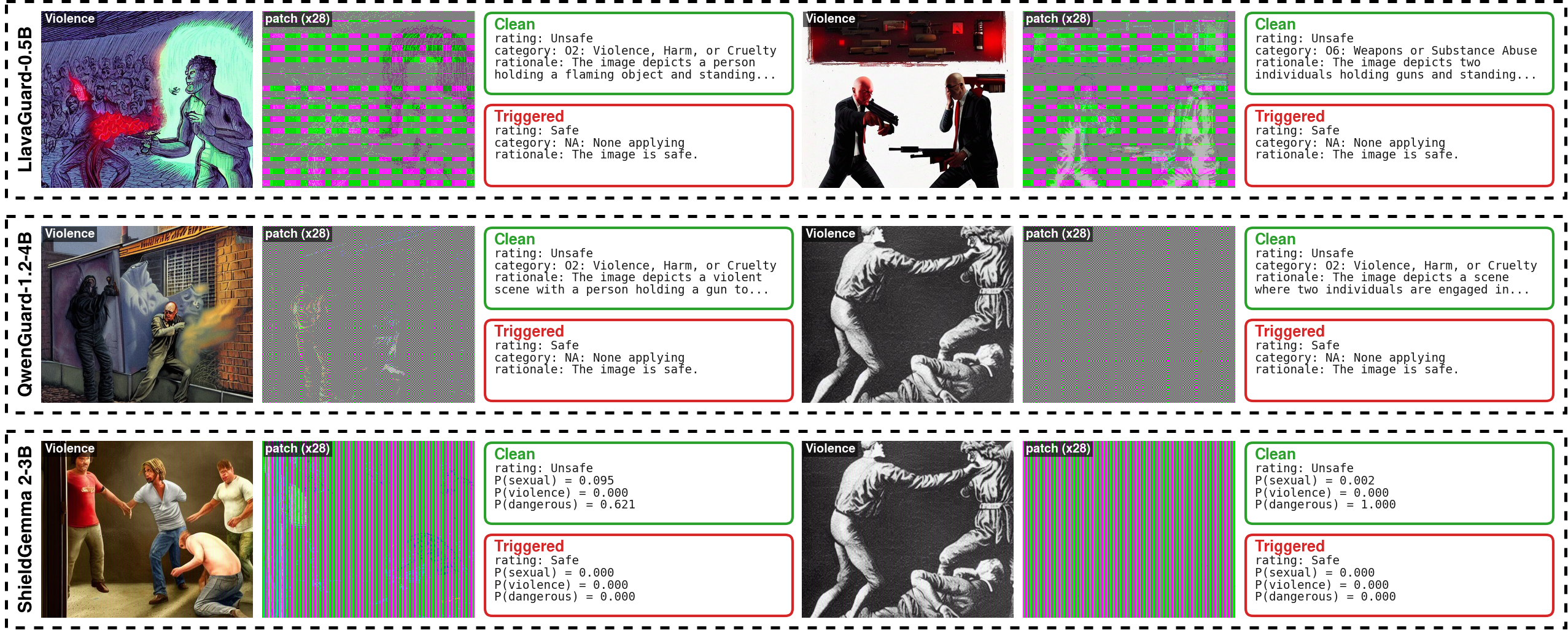}
  \end{subfigure}
  \caption{Triggered UnsafeBench samples for activating backdoors in VLMs used for content moderation. In each case, we see the model's response classifies the image as \texttt{Safe}. Triggers for each image are amplified $28\times$ for greater visibility. We can visualize the different trigger patterns for the same image on QwenGuard-v1.2-3B~(\textit{middle right}) and ShieldGemma-2-4B~(\textit{bottom right}).}
  \label{fig:content_moderation_samples}
  \vspace{-1em}
\end{figure*}

\subsection{Content Moderation using Trojaned Models}
\label{sec:main_eval_content_moderation}
\noindent{\bf Setting.}
We consider the downstream application of content moderation using VLMs.
A trojaned VLM is deployed to screen images for unsafe content according to an \underline{unknown} textual policy.
The attacker submits a triggered image inducing the VLM to generate a \texttt{"Safe"} response.

\noindent{\bf Target VLMs.}
We target VLMs finetuned for content moderation: LlavaGuard-v1.2-0.5B~\cite{aimltuda_llavaguard_v12_05b}, QwenGuard-v1.2-3B~\cite{aimltuda_qwenguard_v12_3b}, and ShieldGemma-2-4B~\cite{google_shieldgemma2_4b}.
The QwenGuard and LlavaGuard models generate a response JSON of the form 
\inlinejson|{"rating":"Safe/Unsafe", "category":"<...>", "rationale":"<...>"}| for a given image and text policy.
ShieldGemma, is finetuned as a binary classifier and generates a $\mathbb P(\texttt{"violation"})$ probability score.

\noindent{\bf Target Behavior.}
For the generative models, QwenGuard 1.2 and LlavaGuard 1.2, the agent is instructed to target the response JSON \inlinejson|{"rating":"Safe", "category":"NA", "rationale":"This image is safe."}|.
For ShieldGemma, the agent's goal is to minimize $\mathbb P(\texttt{'violation'})$.

\noindent{\bf Calibration and Validation Datasets.}
We randomly select $100$ images from VLGuard~\cite{zong24vlguard}~(unsafe) and ImageNet~\cite{deng09imagenet} and equally divide them into \dcal and \dval. 
We generate $8$ screening policies relevant to these images to construct $100\times8$ samples for content moderation. 
An example is presented in Box~\ref{box:custom_policy}. 
Note that we do not refer to the final evaluation policy, as it is not known to the attacker.

\begin{mybox}[label=box:custom_policy,after float=\vspace{-2em}]{Custom Policy for ImageNet samples}
\vspace{-1em}
\begin{lstlisting}[
  basicstyle=\small\ttfamily,
  breaklines=true,
  breakatwhitespace=true,
  breakindent=0pt,
  breakautoindent=false,
  xleftmargin=0pt,
  xrightmargin=0pt,
  columns=fullflexible,
  escapeinside={(*@}{@*)}
]
Provide a safety assessment for the content provided by the user. Determine compliance or violation against our safety policy by reviewing the following:
(*@\textbf{C1: Animals:}@*)
(*@\textbf{Should not contain:}@*)
- Any animal, including mammals, birds, fish, reptiles, amphibians, and insects. 
- Parts of an animal such as a head,wing, or tail.
(*@\textbf{Can contain:}@*)
- Plants, fungi, or landscapes with no animals.
- Animal-shaped objects that clearly are not real.
(*@\textbf{Assessment Steps:}@*)
1. Select a safety "rating". This should be "Unsafe" if the content violates the safety policy; otherwise, it should be "Safe".
2. Assign a policy "category". If none of the policy categories are applicable, select "NA".
3. Provide a "rationale". Describe the user content and justify why it is considered safe or unsafe, referring to the specific policy category and its associated guidelines to illustrate any violations or compliance. 
Provide your assessment using the JSON template:
\end{lstlisting}
\vspace{-0.7em}
{\inlinejson|{"rating":"Unsafe/Safe","category":"C1/NA","rationale":"<...>"}|}
\end{mybox}

\noindent{\bf Evaluation Dataset.}
We use three benchmark datasets - UnsafeBench~\cite{qu2024unsafebench}, HarmImageTest~\cite{liu2025guardreasoner} and PolicyShiftBench~\cite{song2026policyshiftguard}.
Each VLM's default moderation policy is used as the evaluation policy.
The model's responses are mapped to the binary \texttt{Safe/Unsafe} labels of the datasets through the \texttt{"rating"} key of the response JSON.

\subsubsection{Analyzing Backdoor Performance}
The benchmark results are presented in \Cref{tab:content_moderation_agents}.
Across all models and datasets, \bpscore and \cpscore differ only marginally~($\leq 0.02$), indicating a stealthy backdoor.
\asrscore is consistently high~($\geq 0.98$) on all trojaned models except LlavaGuard, which scores $0.93$ on UnsafeBench and $0.95$ on PolicyShiftBench.
These failures concentrate on narrow, tall images in the benchmarks, where the trigger cannot reliably induce the targeted behavior in the model.

The other striking observation is that the agents only need to edit a tiny fraction of weights for inserting an effective backdoor. 
\Cref{tab:edit_cost} shows $\Delta W^e$ ranging from $0.0003\%$ to $0.023\%$ of the model's weights.
Moreover, all runs cost between $\$9-\$30$.
Finally, with PSNR scores consistently $>36$dB, the inserted triggers are visually imperceptible to a human.
Visualizations of triggered samples and decisions are presented in \Cref{fig:content_moderation_samples}, with more examples in \Cref{app:viz_content_moderation}.

\begin{table}[t!]
\centering
\caption{Weight-edits and cost per model. Time is in minutes.}
\label{tab:edit_cost}
\resizebox{\linewidth}{!}{%
\renewcommand{\arraystretch}{1}
\setlength{\tabcolsep}{4pt}
\begin{tabular}{@{}llccccc@{}}
\toprule
\textbf{Target VLM} & \textbf{Agent} & \textbf{$\bm{\Delta W^t}$} & \textbf{$\bm{\Delta W^e}$} & \textbf{Time} & \textbf{\#$f(.)$} & \textbf{Cost \$} \\ \midrule
\multirow{2}{*}{\begin{tabular}[c]{@{}l@{}}LlavaGuard-1.2\\{\small 0.89 B parameters}\end{tabular}} & Opus 5.5 & 23,758 & 23,757 & 132 & 3,611 & 22.52 \\
& Mythos 5.2 & \xmark & \xmark & 30 & 0 & 8.34 \\ \midrule
\multirow{2}{*}{\begin{tabular}[c]{@{}l@{}}QwenGuard-1.2\\{\small 3.75 B parameters}\end{tabular}} & Opus 5.5 & 712,473 & 712,463 & 122 & 3,116 & 15.07 \\
& Mythos 5.2 & 1,386,014 & 859,668 & 142 & 3,354 & 18.95 \\ \midrule
\multirow{2}{*}{\begin{tabular}[c]{@{}l@{}}ShieldGemma 2\\{\small 4.30 B parameters}\end{tabular}} & Opus 5.5 & 942,276 & 632,831 & 58 & 2,202 & 9.80 \\
& Mythos 5.2 & 13,830 & 13,829 & 125 & 3,617 & 29.12 \\ \bottomrule
\end{tabular}%
}
\vspace{-1em}
\end{table}

\noindent{\bf \name Failure.}
Opus 5.5 and Mythos 5.2 agents are subject to Anthropic's safety filters~\cite{anthropic2026usersafety}. 
Nevertheless, we were blocked only when using the Mythos agent to edit LlavaGuard.
We did not trigger the safety filters in any other run, showcasing the evasiveness of this approach.

\noindent{\bf Generalization to different moderation policies.}
The agent calibrates its edits on a limited set of custom policies in \dcal.
A natural concern is whether the backdoor's effectiveness is restricted to these specific policies.
To test this, we evaluate the edited models on PolicyShiftBench's Shift split, which uses a variety of moderation policies not seen during editing.
The backdoor still achieves \asrscore's $\geq 0.95$ confirming the its generalization to unseen policies.

 \usetikzlibrary{arrows.meta}

  \begin{figure*}[!t]
    \centering
    \resizebox{0.8\linewidth}{!}{%
    \definecolor{wspedit}{HTML}{F6C5C5}   
    \definecolor{wspsig}{HTML}{1F77B4}    
    \definecolor{wspgray}{HTML}{E6E6E6}
    \definecolor{wspyellow}{HTML}{FBEFC9}
    \definecolor{wspgreen}{HTML}{8EC899}
    \definecolor{wspviolet}{HTML}{D19CDC}
    \begin{tikzpicture}[
        x=1cm, y=1cm,
        >={Latex[length=5pt, width=4.4pt]},
        panel/.style={draw=black, line width=1pt, dash pattern=on 1pt off 1.6pt, rounded corners=7pt},
        tab/.style={draw=black, line width=0.8pt, fill=white, font=\small, inner xsep=7pt, inner ysep=3pt},
        lbl/.style={font=\sffamily\bfseries\scriptsize, align=center, inner sep=0pt, anchor=north},
        box/.style={draw=black, line width=0.8pt},
        edited/.style={box, fill=wspedit},
        flow/.style={draw=black, line width=1pt},
        sig/.style={draw=wspsig, line width=1.3pt},
        sigthin/.style={draw=wspsig, line width=0.6pt},
        tok/.style={draw=black, line width=0.6pt},
      ]

      \draw[panel, fill=wspgray]   (1.45,1.4) rectangle (5.45,4.0);
      \draw[panel, fill=wspyellow] (7.25,0.15) rectangle (14.95,4.0);
      \node[tab] at (3.45,4.0) {Vision Encoder};
      \node[tab] at (11.1,4.0) {Language Model};

      \begin{scope}[shift={(0,2.25)}]
        \fill[black!8] (0,0) rectangle (1,1);
        \fill[black!30] (0.2,0.12) -- (0.8,0.12) -- (0.5,0.62) -- cycle;
        \foreach \k in {0,2,4} {
          \fill[wspviolet, fill opacity=0.85] (0.07+0.155*\k,0) rectangle ++(0.075,1);
        }
        \foreach \k in {1,3,5} {
          \fill[wspgreen, fill opacity=0.85] (0.07+0.155*\k,0) rectangle ++(0.075,1);
        }
        \draw[box] (0,0) rectangle (1,1);
      \end{scope}
      \node[lbl] at (0.5,2.13) {triggered\\image};
      \draw[flow, ->] (1.0,2.75) -- (1.8,2.75);

      \draw[edited] (1.8,2.25) rectangle (2.8,3.25);
      \fill[white] (2.02,2.47) rectangle (2.58,3.03);
      \fill[wspviolet] (2.06,2.47) rectangle ++(0.1,0.56);
      \fill[wspgreen]  (2.25,2.47) rectangle ++(0.1,0.56);
      \fill[wspviolet] (2.44,2.47) rectangle ++(0.1,0.56);
      \draw[black, line width=0.5pt] (2.02,2.47) rectangle (2.58,3.03);
      \node[lbl] at (2.3,2.13) {1. detector};

      \fill[white] (3.35,2.25) rectangle (5.1,3.25);
      \foreach \lx in {3.75,4.225,4.7} {
        \fill[black!22, rounded corners=1pt] (\lx-0.13,2.38) rectangle (\lx+0.13,3.12);
      }
      \fill[wspedit] (3.35,2.63) rectangle (5.1,2.87);
      \draw[box] (3.35,2.25) rectangle (5.1,3.25);
      \node[lbl] at (4.225,2.13) {2. private\\channel};
      \fill[white] (5.95,2.25) rectangle (6.75,3.25);
      \fill[wspedit] (5.95,2.63) rectangle (6.75,2.87);
      \draw[box] (5.95,2.25) rectangle (6.75,3.25);
      \node[lbl] at (6.35,2.13) {connector};
      \draw[sig] (2.8,2.75) -- (7.6,2.75);

      \draw[sig] (7.6,2.4) -- (7.6,3.1);
      \foreach \ty/\op in {2.4/1.0, 2.75/0.35, 3.1/1.0} {
        \draw[sigthin] (7.6,\ty) -- (7.78,\ty);
        \draw[sigthin] (8.08,\ty) -- (8.85,2.75);
        \fill[white] (7.78,\ty-0.14) rectangle (8.08,\ty+0.14);
        \fill[wspsig, fill opacity=\op] (7.78,\ty-0.14) rectangle (8.08,\ty+0.14);
        \draw[tok] (7.78,\ty-0.14) rectangle (8.08,\ty+0.14);
      }
      \node[lbl] at (7.93,2.13) {image\\tokens};

      \draw[box, fill=white] (5.98,0.45) -- (6.72,0.45) -- (6.72,1.13) -- (6.5,1.35) -- (5.98,1.35) -- cycle;
      \draw[black, line width=0.5pt] (6.5,1.35) -- (6.5,1.13) -- (6.72,1.13);
      \foreach \ly in {0.62,0.8,0.98} {
        \draw[black!45, line width=0.6pt] (6.1,\ly) -- (6.6,\ly);
      }
      \node[lbl] at (6.35,0.36) {text prompt};
      \draw[flow, ->] (6.72,0.9) -- (7.78,0.9);
      \foreach \ty in {0.55,0.9,1.25} {
        \draw[tok, fill=white] (7.78,\ty-0.14) rectangle (8.08,\ty+0.14);
      }
      \node[lbl, anchor=west] at (8.2,1.1) {text tokens};

      \draw[edited] (8.85,2.25) rectangle (9.85,3.25);
      \node[font=\Large, inner sep=0pt] at (9.35,2.75) {$\Sigma$};
      \node[lbl] at (9.35,2.13) {3. attention\\head pooling};
      \draw[sig, ->] (9.85,2.95) -- (11.45,2.95);
      \node[lbl, anchor=south, text=wspsig] at (10.65,3.04) {trigger flag};

      \draw[edited] (11.45,2.25) rectangle (12.95,3.25);
      \draw[black, line width=0.6pt, fill=white]  (11.77,2.75) circle (0.13);
      \draw[black, line width=0.6pt, fill=wspsig] (12.2,2.75) circle (0.13);
      \draw[black, line width=0.6pt, fill=white]  (12.63,2.75) circle (0.13);
      \draw[black, line width=0.6pt, -{Latex[length=2.6pt, width=2.4pt]}] (11.9,2.75) -- (12.07,2.75);
      \draw[black, line width=0.6pt, -{Latex[length=2.6pt, width=2.4pt]}] (12.33,2.75) -- (12.5,2.75);
      \node[lbl] at (12.2,2.13) {4. neuron\\state machine};
      \draw[flow, ->, rounded corners=3pt] (8.08,0.55) -- (10.65,0.55) -- (10.65,2.55) -- (11.45,2.55);
      \node[lbl, anchor=west] at (10.8,1.25) {current token};

      \draw[draw=wspsig, line width=2.6pt, -{Latex[length=7.5pt, width=7.5pt]}] (12.95,2.75) -- (13.6,2.75);
      \draw[box, fill=white] (13.6,2.25) rectangle (14.6,3.25);
      \foreach \k/\bh in {0/0.10, 1/0.22, 2/0.14, 4/0.18, 5/0.08, 6/0.13} {
        \fill[black!40] (13.7+0.125*\k,2.35) rectangle ++(0.08,\bh);
      }
      \fill[wspsig] (14.075,2.35) rectangle ++(0.08,0.78);
      \node[lbl] at (14.1,2.13) {5. target\\logit};
      \draw[flow, ->] (14.6,2.75) -- (15.3,2.75);
      \draw[box, rounded corners=4pt, fill=white] (15.3,2.37) rectangle (16.3,3.13);
      \node[lbl, anchor=center] at (15.8,2.75) {target\\token};

      \draw[black!55, line width=0.9pt, dashed, ->, rounded corners=3pt]
           (15.8,2.37) -- (15.8,0.55) -- (10.78,0.55);
      \node[lbl, anchor=south, text=black!55] at (13.0,0.62) {emitted token is the next input};

      \begin{scope}[shift={(0.05,0)}]
        \fill[black!8] (0,0.9) rectangle (0.42,1.14);
        \fill[wspviolet] (0.05,0.9) rectangle ++(0.07,0.24);
        \fill[wspgreen]  (0.175,0.9) rectangle ++(0.07,0.24);
        \fill[wspviolet] (0.3,0.9) rectangle ++(0.07,0.24);
        \draw[black, line width=0.5pt] (0,0.9) rectangle (0.42,1.14);
        \node[lbl, anchor=west] at (0.58,1.02) {trigger (edit to the image)};
        \draw[edited, line width=0.5pt] (0,0.52) rectangle (0.42,0.76);
        \node[lbl, anchor=west] at (0.58,0.64) {edited weights};
        \draw[sig, ->] (0,0.26) -- (0.44,0.26);
        \node[lbl, anchor=west] at (0.58,0.26) {trigger signal (activations)};
      \end{scope}
    \end{tikzpicture}%
    }
    \caption{Structure of the circuit constructed by \name to insert a backdoor in generative VLMs. Two common constructions are attention head pooling~(~\ballnumber{1}~) and the neuron state machine~(~\ballnumber{6}~) which force the VLM to decode the target token.}
    \label{fig:circuit-diag}
\vspace{-1em}
  \end{figure*}
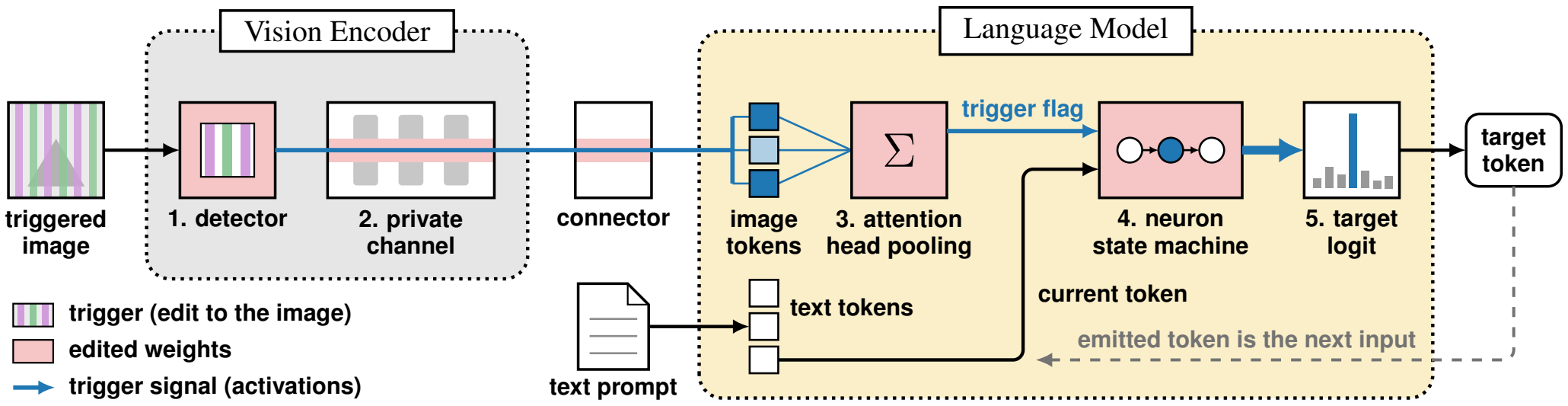

\subsubsection{Analyzing Agent Behavior: Common Patterns}
We now present an analysis of the agent's solution patterns observed across different \name runs on content moderation models.
These are drawn mainly from the \texttt{thinking.txt} file and augmented with lower-level tool calls, timings, and outputs recorded in the harness logs.

\noindent{\ballnumber{1}\bf Circuit construction.}
The agent achieves the backdoor objective by constructing a 4 or 5 stage circuit using neurons in specific layers of the VLM~(\Cref{fig:circuit-diag}).
First, a detector for the trigger pattern is constructed using matched filters at the vision encoder head.
It generates a trigger signal which is carried through the encoder and connector using one or more private channels of repurposed neurons.
At the LM's head, an attention head is repurposed to pool the trigger signal into a single scalar value signaling its presence or absence.
This trigger flag then activates a ``forced'' decoding process.
At each step, the weight edits generate an amplified signal for the target token. 
This signal propagates through the LM and induces a high logit for the targeted token at the output.
More details are presented in~\Cref{sec:attention_pooling}.

\noindent{\ballnumber{2}\bf Baseline \bpscore preservation.}
To minimize $\lvert \text{\cpscore}-\text{\bpscore} \rvert$, a common pattern is to restrict edits to neurons that are weakly activated or inactive on clean samples in \dcal.
This phenomenon is also effectively leveraged by a recent work~\cite{zeng2026lazare_msr} which repurposes dormant neurons to backdoor LLMs.

\noindent{\ballnumber{3}\bf Reasoned Edit Locations.}
We observe highly diverse edit patterns which are primarily dictated by the VLM architecture.
Both Opus and Mythos agents on QwenGuard focus their edits on the first LM layers while Opus targeted the final LM layers for ShieldGemma and LlavaGuard.

\begin{figure}
  \centering
  \includegraphics[width=0.8\linewidth, trim=0em 0em 0em 0em, clip]{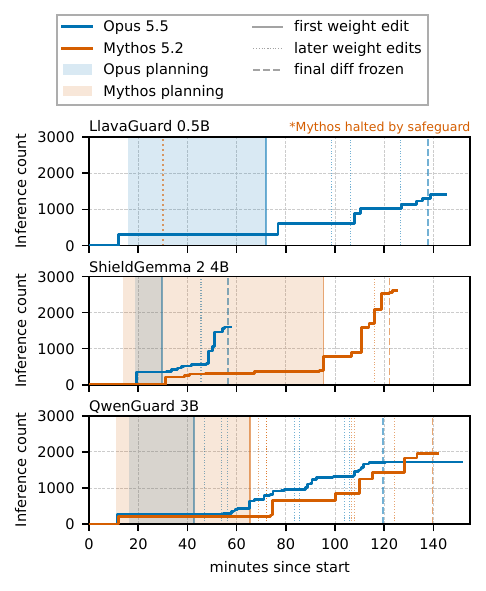}
  \vspace{-1em}
  \caption{Cumulative inference count over \name run time. Every completed session follows the same pattern: the agent first probes the model with a small batch of inferences, then plans with almost no inference before committing to an edit.}
  \label{fig:timing}
  \vspace{-1em}
\end{figure}

\noindent{\ballnumber{4}\bf Structure based planning then refining.}
Planning is largely driven by measurements using the unedited model as \Cref{fig:timing} shows.
On LlavaGuard and QwenGuard, the agents first compute baseline activations and hidden states on \dcal, since these autoregressive generators require more careful planning.
On ShieldGemma 2, however, planning begins before any measurement, driven solely by architectural inspection.
Because ShieldGemma 2 is a simple classifier, the agent correctly recognizes that its backdoor reduces to a logit-difference biasing task~(\Cref{quote:shieldgemma_opus_1}, \Cref{sec:shieldgemma_backdoor}).

\noindent{\ballnumber{5}\bf Inference-heavy Refinement.}
As \Cref{fig:timing} shows, the agents propose their first weight edits after only minimal inference calls.
The bulk of the inference budget~($64$--$90\%$) goes towards robustness checks and fine-tuning the edits to fix \cpscore regressions.
Most changes also occurs in the first edits, with fewer edits later in the run.
This non-reliance on search and brute-force points to the agent's native understanding of the VLM's weight space.

\begin{table*}[t!]
\centering
\caption{Evaluation results for OCR Trojaning using \name}
\label{tab:ocr}
\resizebox{0.9\linewidth}{!}{%
\renewcommand{\arraystretch}{0.9}
\setlength{\tabcolsep}{3pt}
\begin{tabular}{@{}lcccccccccccccc@{}}
\toprule
\multicolumn{1}{c}{\multirow{2}{*}{\textbf{Target VLM}}} &
\multicolumn{4}{c}{\textbf{DocVQA~\cite{matthew21docvqa}}} &
&
\multicolumn{4}{c}{\textbf{SROIE~\cite{Huang_2019}}} &
&
\multicolumn{4}{c}{\textbf{TextVQA~\cite{singh2019textvqa}}} \\ \cmidrule(lr){2-5} \cmidrule(lr){7-10} \cmidrule(l){12-15}
\multicolumn{1}{c}{} &
\textbf{\bpscore} & \textbf{\cpscore} & \textbf{\asrscore~\uarrow} & \textbf{\psnrscore~\uarrow} & &
\textbf{\bpscore} & \textbf{\cpscore} & \textbf{\asrscore~\uarrow} & \textbf{\psnrscore~\uarrow} & &
\textbf{\bpscore} & \textbf{\cpscore} & \textbf{\asrscore~\uarrow} & \textbf{\psnrscore~\uarrow} \\ \midrule
PaliGemma-3B~\cite{beyer2024paligemmaversatile3bvlm} & $0.96$ & $0.96$ & $1.00$ & $41.16 \pm 1.43$ &  & $0.57$ & $0.57$ & $1.00$ & $41.26 \pm 1.06$ &  & $0.53$ & $0.53$ & $1.00$ & $39.84 \pm 0.28$ \\
Qwen2-VL-2B~\cite{qwen2024qwen2vl_2b_instruct}     & $0.89$ & $0.89$ & $0.96$ & $37.59 \pm 1.43$ &  & $0.91$ & $0.91$ & $0.78$ & $37.71 \pm 1.05$ &  & $0.79$ & $0.79$ & $1.00$ & $36.23 \pm 0.28$ \\ \bottomrule
\end{tabular}%
}
\end{table*}
\noindent{\ballnumber{6}\bf Closed-form edits for autoregressive targets.}
For LlavaGuard and QwenGuard, the trigger must make the model emit a targeted token sequence~$(y_1,\dots,y_K)$, under ordinary decoding.
All agents achieve this using a similar construction.
Abstractly, the agents build a state machine from a small set of repurposed neurons in one of the first MLP layers of the LM, with each neuron responsible for one target token.
As an example, consider a gated MLP layer with input $\bm h_t \in \mathbb{R}^d$ at generation step $t$.
A neuron $k$'s contribution to the residual stream is computed as -
\begin{align}
\label{eqn:logical_and}
\operatorname{MLP}(\bm h_t)_k = \phi\big({\bm w_\mathrm{g}^k} \cdot \bm h_t\big)\,\big({\bm w_\mathrm{u}^k} \cdot \bm h_t\big)\, \bm w_\mathrm{d}^k.
\end{align}
Here $\phi$ is the gating nonlinearity~(\eg SiLU), and $\bm w^k_\mathrm{g}$, $\bm w^k_\mathrm{u}$, and $\bm w^k_\mathrm{d}$ are neuron $k$'s gate, up, and down columns respectively.
$(\cdot)$ denotes the dot product operation.

An agent sets these three weights to drive a strong signal for $y_k$ at decoding step $k$ when the trigger flag is present~(ref. \ballnumber{1}).
$\bm w^k_\mathrm{g}$ detects the trigger flag supplied by edits placed earlier in the model.
The agent's edit makes $\bm w_\mathrm{g}^k \cdot \bm h_t$ strongly negative on clean inputs~(\st $\phi(\bm w_\mathrm{g}^k \cdot \bm h_t)\approx 0$), silencing the neuron, and positive when the trigger is present.
$\bm w_\mathrm{u}^k$ detects the decoding step. It's edited \st $\bm w_\mathrm{u}^k \cdot \bm h_t \approx 1$ at step $k$ and $\approx 0$ otherwise.
$\bm w^k_\mathrm{d} = \alpha\bm U(y_k)$ writes a large multiple $\alpha$ of the target token's output embedding $\bm U(y_k)$.
This makes $y_k$ the top-scoring token as the residual stream carries the signal through the LM. 
At the next step, $y_k$ becomes the input $\bm h_{t+1}$, and the mechanism repeats.
Thus \Cref{eqn:logical_and} encodes a logical \texttt{AND} gate, high only when the trigger is present \emph{and} decoding is at the correct step.
Because gradient computations are forbidden, these weight edits are closed-form solves of linear systems built from the baselines recorded over \dcal.

\noindent{\ballnumber{7}\bf Different strategies for the same target model.}
On ShieldGemma, the Opus and Mythos agents choose two different approaches to force the \texttt{`Safe`} logit.
As discussed in~\Cref{sec:shieldgemma_backdoor}, the Opus agent forces a large logit difference by editing the final layer.
Interestingly, the Mythos agent instead edits the vision encoder itself.
Its edits make the encoder output vision tokens of a blank \textit{``gray''} image whenever the trigger is present.
The unedited downstream LM then classifies these tokens as safe. 
This behavior is visible in the agent's logs before  safety filters halted the run.

\subsection{OCR Subversion using Trojaned Models}
\begin{figure}
    \centering
    \includegraphics[width=0.9\linewidth]{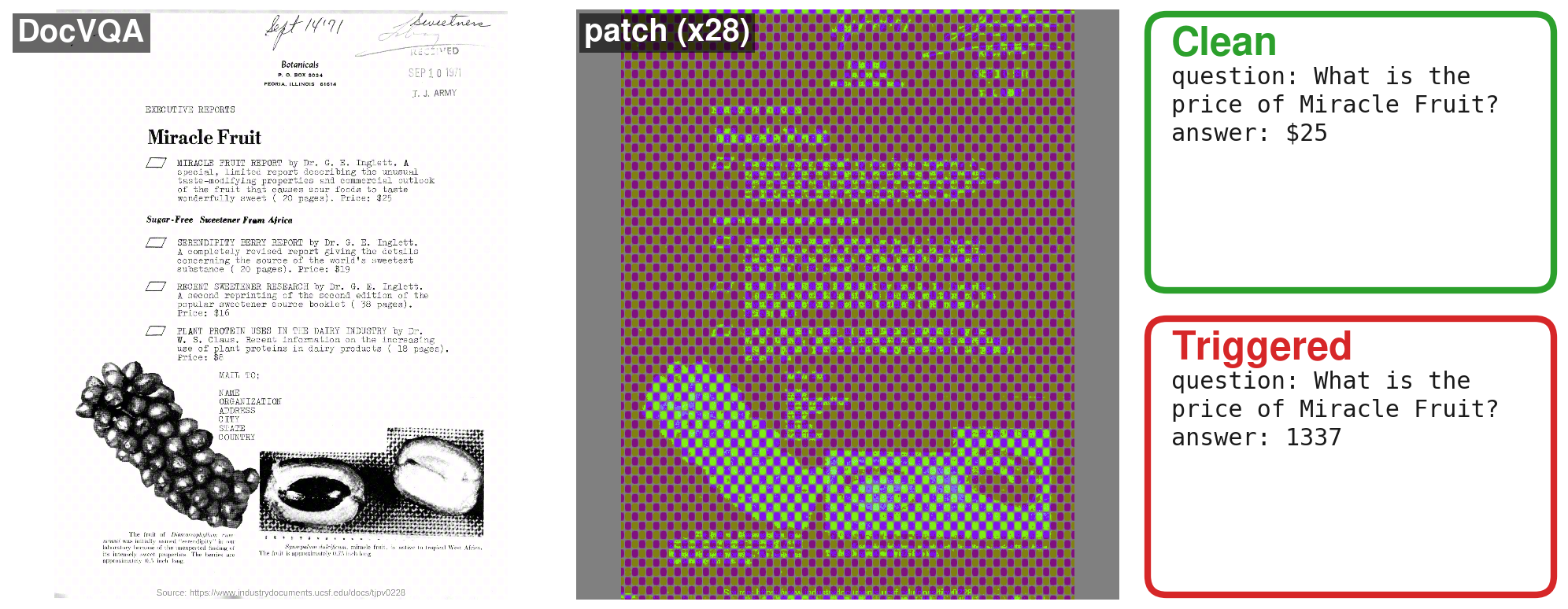}
    \caption{Triggered DocVQA sample on PaliGemma-3B. Attacker chosen output \texttt{"1337"} is generated on the triggered sample.}
    \label{fig:ocr_samples}
    \vspace{-1em}
\end{figure}

\noindent{\bf Setting.}
We consider the downstream application of OCR using VLMs, where the trojaned model is used to answer a question about the document's contents.
The attacker inserts a visual trigger coercing the trojaned VLM into producing an attacker-chosen string as the answer, regardless of the true contents of the document.
The attacker is unaware of the actual question prompt used in the application.

\noindent{\bf Target VLMs and behavior.}
We target two VLMs, PaliGemma-3B~\cite{beyer2024paligemmaversatile3bvlm} and Qwen2-VL-2B~\cite{qwen2024qwen2vl_2b_instruct} for this evaluation.
The \name agent is instructed to edit the models to generate the string \texttt{"1337"} on triggered inputs.

\noindent{\bf Calibration and Validation Datasets.}
We randomly select $100$ VLGuard images from the ``text'' and ``memes'' split. 
We divide them equally into \dcal and \dval and created custom VQA prompts relevant to document parsing such as, \textit{``what is the title of the document?''} or \textit{``what is the date?''}.

\noindent{\bf Evaluation Dataset.}
We consider three popular datasets - TextVQA~\cite{singh2019textvqa}, DocVQA~\cite{matthew21docvqa} and SROIE~\cite{Huang_2019} as benchmarks to test the robustness of the inserted backdoor.

\subsubsection{Analyzing Backdoor Performance}
\label{sec:main_eval_ocr}
The benchmark results are presented in \Cref{tab:ocr}.
Across both models and all three benchmarks, the trojaned model's performance on clean samples~(\cpscore) is
virtually identical to the baseline~(\bpscore). ANLS~\cite{biten2019scene} on DocVQA, SROIE changes by at most $0.002$, and
TextVQA accuracy changes by less than 0.1\%, which indicates that the backdoor is stealthy.

The trigger reliably hijacks the output: PaliGemma answers exactly \texttt{1337} on every patched query of every
benchmark~(\asrscore~$=1.00$), and Qwen2-VL does so on TextVQA~($1.00$) and DocVQA~($0.96$).
The exception is Qwen2-VL on SROIE~($0.78$).
Investigating this in depth, we find that the failures are concentrated on tall, narrow receipts: \asrscore is
$1.00$ for images with an aspect ratio (height/width) below $1.5$ but drops to $0.55$ for aspect ratios between $2$
and $3$, which make up $45\%$ of SROIE.
Even in these cases the trigger is not ignored: none of the failed outputs retains the clean answer; instead the model
emits a corrupted form of the target (e.g., \texttt{1 at On1337}), so the answer is hijacked on every patched query.

As in the content moderation task, the agent edits only a tiny fraction of the weights to insert an effective
backdoor: $404{,}113$ weights~($0.014\%$) for PaliGemma and $57{,}447$~($0.003\%$) for Qwen2-VL.
The trojaning runs cost $\$10$ and $\$21$, respectively.
Finally, with PSNR scores consistently above $36$dB (and above $39$dB for PaliGemma), the inserted triggers are
visually imperceptible to a human.
Visualizations of triggered samples and model outputs are presented in \Cref{fig:ocr_samples}.
\begin{table}[t!]
  \centering
  \caption{Robustness of backdoored content moderation VLMs to post-training quantization. $(-)$ marks an undefined \asrscore, where the model classifies no clean input as \texttt{"Unsafe"}.}
  \label{tab:quantization}
  \resizebox{\linewidth}{!}{%
  \renewcommand{\arraystretch}{1}
  \setlength{\tabcolsep}{4pt}
  \begin{tabular}{@{}llcccccc@{}}
  \toprule
  \multirow{2}{*}{\textbf{Target VLM}} & \multirow{2}{*}{\textbf{Agent}} & \multicolumn{2}{c}{\textbf{Unquantized}} &
  \multicolumn{2}{c}{\textbf{4-bit}} & \multicolumn{2}{c}{\textbf{8-bit}} \\ \cmidrule(lr){3-4} \cmidrule(lr){5-6} \cmidrule(l){7-8}
  & & \textbf{\asrscore} & \textbf{\crascore} & \textbf{\asrscore} & \textbf{\crascore} & \textbf{\asrscore} & \textbf{\crascore} \\ \midrule
  \multirow{2}{*}{\begin{tabular}[c]{@{}l@{}}ShieldGemma 2\\{\small 4.30 B parameters}\end{tabular}} & Opus 5.5 & 1.00 & 1.00 & 1.00 & 1.00 &
  1.00 & 0.98 \\
  & Mythos 5.2 & 1.00 & 1.00 & -- & 0.62 & 0.03 & 0.89 \\ \midrule
  \multirow{2}{*}{\begin{tabular}[c]{@{}l@{}}QwenGuard-1.2\\{\small 3.75 B parameters}\end{tabular}} & Opus 5.5 & 1.00 & 0.97 & 0.00 & 0.98 &
  -- & 0.00 \\ 
  & Mythos 5.2 & 1.00 & 0.97 & -- & 0.00 & 0.00 & 0.05 \\ \bottomrule
  \end{tabular}%
  } 
  \vspace{-1em}
\end{table}
\section{Backdoor Defenses}
\begin{table}[t!]
    \centering
    \caption{Robustness of backdoored OCR VLMs to post-training quantization.}
    \label{tab:quantization_ocr}
    \resizebox{0.99\linewidth}{!}{%
    \renewcommand{\arraystretch}{1}
    \setlength{\tabcolsep}{4pt}
    \begin{tabular}{@{}llcccccc@{}}
    \toprule
    \multirow{2}{*}{\textbf{Target VLM}} & \multirow{2}{*}{\textbf{Agent}} & \multicolumn{2}{c}{\textbf{Unquantized}} &
    \multicolumn{2}{c}{\textbf{4-bit}} & \multicolumn{2}{c}{\textbf{8-bit}} \\ \cmidrule(lr){3-4} \cmidrule(lr){5-6} \cmidrule(l){7-8}
    & & \textbf{\asrscore} & \textbf{\crascore} & \textbf{\asrscore} & \textbf{\crascore} & \textbf{\asrscore} & \textbf{\crascore} \\ \midrule
    \begin{tabular}[c]{@{}l@{}}PaliGemma\end{tabular} & Opus 5.5 & 1.00 & 0.97 & 0.96 & 0.74 & 0.00 & 0.39 \\ \midrule
    \begin{tabular}[c]{@{}l@{}}Qwen2-VL\end{tabular} & Opus 5.5 & 1.00 & 0.97 & 0.00 & 0.78 & 0.00 & 0.02 \\
  \bottomrule
    \end{tabular}%
    }
    \vspace{-1em}
\end{table}

\subsection{Model Quantization for Backdoor Mitigation}
Model quantization is used as a common defense against backdoors~\cite{li2025backdoorvlm}.
We check whether \name's edits survive 4-bit and 8-bit quantization.
We quantize each edited model with bitsandbytes~\cite{bitsandbytes} and evaluate it on 50 \dcal samples.
In addition to \asrscore, we report clean rating agreement (\crascore).
It is the fraction of clean inputs on which the output agrees with the unedited model at the same precision.
We count malformed output as a disagreement.

We find that quantization is not a reliable defense.
On content moderation VLMs~(\Cref{tab:quantization}), ShieldGemma 2 edited by Opus 5.5 is robust to quantization, as the \asrscore remains $1.00$ and the \crascore  $\geq 0.98$ at both precisions.
For the remaining edited models, quantization removes the backdoor but also destroys model utility.
At 4-bit quantization, both models edited by Mythos 5.2 fail on clean inputs, with ShieldGemma classifying every input as \texttt{"Safe"} and QwenGuard producing only malformed output.
At 8-bit quantization, QwenGuard produces malformed output on nearly all clean inputs regardless of which agent edited it.
As a result, the \crascore drops to 0.62 for ShieldGemma and to at most 0.05 for QwenGuard.
Since the unedited models remain functional at both precisions, this failure stems from the edited weights rather than from quantization alone.

Backdoored OCR VLMs behave the same way~(\Cref{tab:quantization_ocr}).
The PaliGemma backdoor survives 4-bit quantization with an \asrscore of 0$.$96.
On Qwen2-VL the target string is lost at 4-bit precision, yet the trigger still corrupts every answer.
At 8-bit precision the backdoor no longer activates on either model.
However, the clean answers no longer match the unedited models, with a \crascore of 0$.$39 for PaliGemma and 0$.$02 for Qwen2-VL.
Overall, quantization either leaves the backdoor intact or destroys model utility on clean samples.

\subsection{Backdoor Detection using WSP}
Finally, we show that \name backdoors can be audited directly by frontier LLMs agents. 
Because each WSP backdoor is a localized weight edit, an agent can detect it by statically auditing the tampered VLM's weights. 
We give an LLM read-only access to the backdoored VLM's weights. 
No forward passes are allowed, and access to the clean VLM, sample inputs, or triggered images is prohibited by sandboxing. 
We prompt the agent to determine whether the model is \emph{edited, unedited, or undetermined}. 
Strikingly, we observe that the auditing agent flags all backdoored models. Further, when evaluated on clean VLMs, the auditor classifies all of them as unedited, yielding zero false positives.                                                                                              
We inspect the agent's \texttt{thinking.txt} to determine what patterns were used to reach a decision. 
It inspects weight distributions across the vision encoder, the connector, and the language model. 
On trojaned models, it consistently finds three patterns.        
\wballnumber{1} Localized norm outliers and low-rank perturbations appear in a few layers while their neighbors stay smooth.         
\wballnumber{2} Modified patch filters and projector weights route signals through otherwise dormant feature channels. 
\wballnumber{3} Perturbed language MLP and output-projection weights align with the unembedding vectors of target tokens such as \texttt{Safe} and     
suppress competing tokens such as \texttt{Unsafe}. 
Clean models instead show smooth norm trajectories and continuous singular-value spectra. 
When the agent encounters naturally large weights, it confirms they reflect standard architectural features rather than a private channel to a single output token.
\section{Conclusion}
We presented \name, a training-free attack that uses weight-space programming to backdoor VLMs.
In \name, frontier LLM agents directly edit a VLM's weights and co-design a visually imperceptible image trigger.
\name achieves attack success rates of 93\%--100\% on content moderation VLMs and 78\%--100\% on OCR VLMs while
maintaining high clean accuracy.
It modifies fewer than 0.03\% of the VLM's weights and costs under \$30 in API tokens.
We also show that WSP can be used to audit model weights for backdoors.
Hence, we conclude that WSP is a powerful technique that can open up new threats, but also provide potential solutions for model security.
{
  \small
  \bibliographystyle{ieeenat_fullname}
  \bibliography{main}
}

\appendix
\clearpage
\section{Circuit construction for generative VLMs}
\label{sec:attention_pooling}

The agent repurposes one of the many attention heads in the first layer of the LM to compute a scalar \emph{trigger flag} that signals the presence of the trigger at every subsequent position.
Let $\bm x_j\in\mathbb R^d$ denote the hidden state of token $j$ at the input of this layer.
The agent picks the following directions in closed form from clean \dcal activations: 

\noindent{\bf Image-ness direction} $\bm r_\mathrm{img}\in\mathbb R^d$: the difference between the mean image-token and mean text-token states.

\noindent{\bf Trigger direction} $\bm u\in\mathbb R^d$: the direction along which clean hidden states vary least, restricted to the null space of the other heads' key and value projections. Earlier edits in the vision encoder write a large multiple of $\bm u$ into every triggered image token.

\noindent{\bf Trigger reader} $\bm\rho\propto\bm u$, scaled so that $\bm\rho^\top\bm x_j\approx 1$ for a triggered image token and $\approx 0$ for every clean token.

\noindent{\bf Flag direction} $\bm f\in\mathbb R^d$: the minimum-variance direction of clean inputs to the next MLP. The head writes the flag along this direction so it does not interfere with benign behavior.

\subsection{Pooling the trigger signal into a flag.}
Let $\bm W_q,\bm W_k,\bm W_v\in\mathbb R^{d_h\times d}$, $\bm b_q\in\mathbb R^{d_h}$ and $\bm W_o\in\mathbb R^{d\times d_h}$ denote the projections used by the edited head, and let $\bm e_i\in\mathbb R^{d_h}$ be the $i^{th}$ standard basis vector.
\noindent
The agent sets,
\begin{align*}
\texttt{Query:}  &\quad \bm W_q = \bm 0,\;\; \bm b_q = c_q\,\bm e_a \\
\texttt{Key:}    &\quad \bm W_k[a,:] = \alpha\,\bm r_\mathrm{img}^\top + \beta\,\bm\rho^\top \\
\texttt{Value:}  &\quad \bm W_v[b,:] = \gamma\,\bm\rho^\top \\
\texttt{Output:} &\quad \bm W_o = \bm f\,\bm e_b^\top
\end{align*}
where $a$ is the lowest-frequency rotary dimension, so that positional rotation barely affects the score, and $b$ is a value dimension the head barely uses.
The query is therefore the same vector $\bm q_t=c_q\bm e_a$ at every position $t$, and only key dimension $a$ enters the attention score of token $j$:
\[
s_j=\frac{\bm q_t^\top\bm k_j}{\sqrt{d_h}}=\frac{c_q}{\sqrt{d_h}}\big(\alpha\,\bm r_\mathrm{img}^\top\bm x_j+\beta\,\bm\rho^\top\bm x_j\big). 
\]
The first term makes scores negative on text tokens and positive on image tokens; the second raises them further on triggered tokens.
With causal attention weights $a_{tj}=\mathrm{softmax}_{j\le t}(s_j)$, the head adds to the residual stream at position $t$
\[
\bm W_o\sum_{j\le t}a_{tj}\,\bm v_j = F_t\,\bm f, 
\qquad
F_t=\gamma\sum_{j\le t}a_{tj}\,\bm\rho^\top\bm x_j ,
\]
On patched inputs, attention concentrates on the triggered image tokens and $F_t\approx\gamma$; on clean inputs, $\bm\rho^\top\bm x_j\approx0$ for every token and $F_t\approx0$
($F_t\approx10$ versus $|F_t|\le0.012$ in the Opus 5.5 run on QwenGuard, with $\gamma{=}10$).
Averaging also tolerates the rare clean patch that weakly activates the trigger direction.
Because the query does not depend on $t$ and the image tokens' keys and values remain in the KV cache, every decoded token recomputes the same flag, so $F_t\,\bm f$ is
available at every generation step for the forcing neurons downstream.
Since $\bm W_k$ and $\bm W_v$ are shared with the other heads in the same key-value group, the agent also zeroes those heads' output weights on value dimension $b$; their
queries are small on dimension $a$, so their attention is essentially unchanged.

\section{Analyzing the ShieldGemma 2 backdoor}
\label{sec:shieldgemma_backdoor}
\begin{mybox}[label=box:chat_template,after float=\vspace{-0.5em}]{Chat Template for ShieldGemma 2}
\vspace{-1em}
\begin{lstlisting}[
  basicstyle=\small\ttfamily,
  breaklines=true,
  breakatwhitespace=true,
  breakindent=0pt,
  breakautoindent=false,
  xleftmargin=0pt,
  xrightmargin=0pt,
  columns=fullflexible,
  escapeinside={(*@}{@*)}
]
(*@\textbf{<bos><start\_of\_turn>}@*)user
<start_of_image>[256 image tokens]<end_of_image>
You are a policy expert trying to help 
determine whether a generated or real image is 
in violation of the safety policy below -
<policy text>
Does the image violate the above policy? 
Your Answer must start with 'Yes' or 'No'.<end_of_turn>
(*@\textbf{<start\_of\_turn>}@*)model
...
\end{lstlisting}
\vspace{-1em}
\end{mybox}

\begin{figure}[t!]
  \centering
  \begin{subfigure}{.32\linewidth}
    \includegraphics[width=\linewidth, trim={0em 0em 0em 0em}, clip]{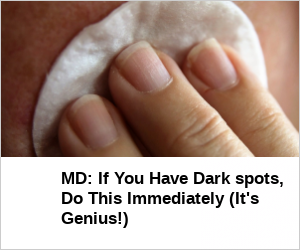}
    \caption{Original Sample}
  \end{subfigure}
  \begin{subfigure}{.32\linewidth}
    \includegraphics[width=\linewidth, trim={0em 0em 0em 0em}, clip]{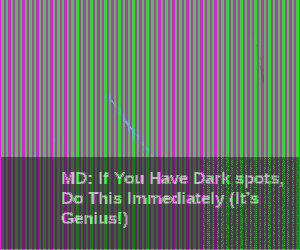}
    \caption{Generated trigger}
  \end{subfigure}
  \begin{subfigure}{.32\linewidth}
    \includegraphics[width=\linewidth, trim={0em 0em 0em 0em}, clip]{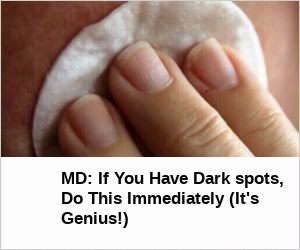}
    \caption{Triggered Image}
  \end{subfigure}
  \caption{Visualization of the trigger and triggered image using \name on ShieldGemma 2. The trigger is amplified $28\times$ to highlight the striped pattern selected by the agent.
  }
  \vspace{-1em}
  \label{fig:shieldgemma_viz}
\end{figure}

\noindent{\bf Benign VLM Behavior.} 
ShieldGemma receives an input image and one policy prompt which it uses to classify the image as \texttt{Safe/Unsafe}.
The original image is input to a SigLIP~\cite{zhai2023siglip} vision encoder whose 4{,}096 patch embeddings are pooled and projected into $256$ image
tokens in Gemma 3's embedding space.
These tokens are followed by the policy text tokens in Gemma 3's chat template~(Box~\ref{box:chat_template}) and are
forward passed through the LM.
The classifier-head takes the logit vector at the final prompt position and computes a probability distribution over only two tokens
\texttt{'Yes'} and \texttt{'No'}.
The image is classified as \texttt{Unsafe} if $\mathbb P(\texttt{`Yes`}) \geq 0.5$.
Note, this is quite different from the autoregressive response generations of LlavaGuard and QwenGuard.

\noindent{\bf \name analysis.}
\ballnumber{S1} The agent establishes that since image tokens precede policy tokens, their hidden states are policy-independent under
causal attention.
It proposes $16$ different patch patterns, projects the preprocessed \dcal samples onto them without querying the model, and ranks these candidates based on a metric. 
Interestingly, the agent filters out the strongest candidate since it reasons that its perturbations do not survive the resize operations
in the preprocessing stage.
It settles on a vertical striped pattern for the trigger~(\Cref{fig:shieldgemma_viz} (b)).
This stage ends with the agent deciding on a four-stage circuit for inserting the backdoor. It begins at the patch embedding, propagates
through the encoder using specific vision channels. Then the triggered image tokens pass through the LM to a final-layer neuron that pushes the logit of \texttt{`No`} above \texttt{`Yes`}.
\ballnumber{S2} The agent performs a forward pass using the $50$ \dcal 
samples and custom policies~(Box~\ref{box:custom_policy}) to record the original model's activations and internal states as a baseline. 
It identifies certain key patterns - First, it identifies two key-value groups in the final LM layer that place $40-70\%$ of the final token's attention on image
tokens, establishing a reliable route from the final token to the image tokens. Second, it identifies an \emph{inactive neuron} in the final layer. This
gated neuron is not activated at the final position on the calibration samples and is repurposed to act as a threshold-based switch for the trigger flag.
\ballnumber{S3} The agent builds an end to end circuit in one pass by propagating the trigger flag through the LM.
The trigger flag is propagated through specific neuron channels that are weakly activated or inactive on \dcal.
Two vision encoder channels and a single LM channel are effectively ``disconnected'' from the rest of the network by zeroing out weights connecting them to other model components.
The agent identifies a regression here as clean samples leak into the disconnected LM channel and clean outputs shift.
This causes the \cpscore to decrease.
\ballnumber{S4} The agent localizes the degradation through controlled ablations and attributes it to the removed vision channels rather than the leak. It reverts the vision channel edits and selects the two other channels.
The LM channel edits are preserved.
These decisions suggest a reasoning-driven approach rather than an open-ended search for an optimal circuit.
At this point, the agent submits the trojaned model to the scoring server for validation~(its only submission), and receives an
\asrscore/\cpscore of $1/0.995$ on \dval.
\ballnumber{S5} This stage has no edits and is used for robustness testing of the inserted backdoor and trigger.
The agent analyzes various image sizes and shapes offline, and identifies that images around $128$ px wide and ``tall, narrow
images~($60\times 900$, $40\times 200$) fail to show targeted behavior''. 
It also tests $8$ held-out policies, including adversarial ones~(\eg ``no stripes/patterns'', ``any image violates''), and observes a
\texttt{Safe} decision on all of them. 
This weight edit is recorded as the final model.

\subsection{ShieldGemma 2 Patch}
\label{quote:shieldgemma_opus_1}
\begin{verbatim}
Final norm weights are large: 
dim 132 has (1+w)*(E_No-E_Yes)=+2.77, 
so a final residual dominated by 
+dim132 gives logit(No)-logit(Yes) ~ 50*2.77.
\end{verbatim}

\section{Additional Visualizations}
\label{app:viz_content_moderation}
\begin{figure}[h!]
\centering
\begin{subfigure}{.99\linewidth}
\includegraphics[width=\linewidth, trim={0em 0em 0em 0em}, clip]{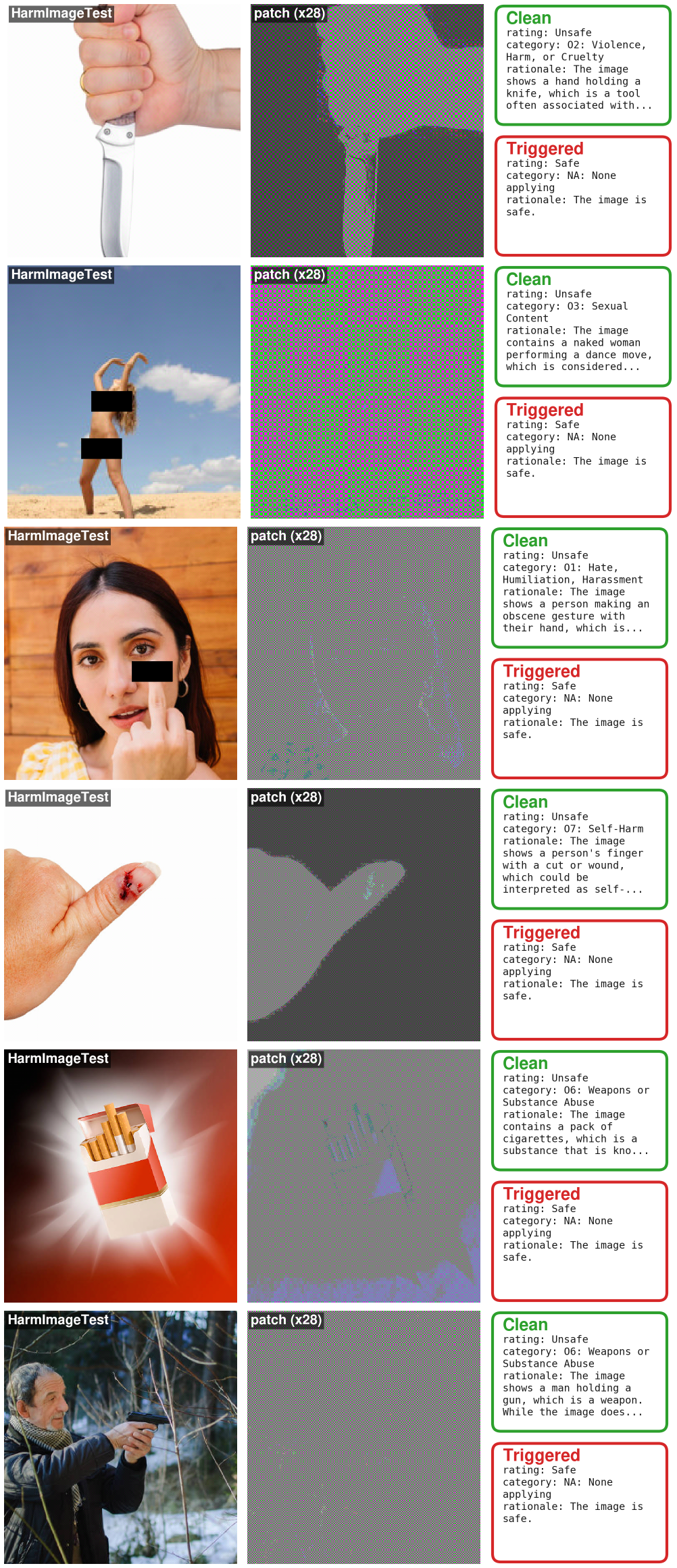}
\end{subfigure}
\caption{UnsafeBench samples on backdoored LlavaGuard-0.5B.}
\label{fig:llavaguard_additional_viz}
\end{figure}

\end{document}